\documentclass[12pt,english,floatfix,nofootinbib,superscriptaddress,aps,prd,preprint]{revtex4}
\usepackage[utf8]{inputenc}
\usepackage{float}
\usepackage{array}
\usepackage{lipsum}
\usepackage{graphicx}
\usepackage{amsmath,amsthm,amsfonts,amssymb}
\usepackage{graphicx}
\usepackage[english]{babel}
\usepackage{color}
\usepackage{subfig}
\usepackage{caption}
\usepackage{tensor}
\usepackage{esint}
\usepackage[dvips]{epsfig}
\usepackage[dvips]{graphicx}
\usepackage{float}
\usepackage{units}
\usepackage{textcomp}
\usepackage{mathrsfs}
\usepackage{amsmath}
\usepackage[makeroom]{cancel}
\usepackage{amssymb}
\usepackage{amsbsy}
\usepackage{amsfonts}
\usepackage{amssymb,mathrsfs,xcolor}
\usepackage{esint}
\usepackage{array}
\usepackage{graphicx}

\usepackage{hyperref}
\hypersetup{
    colorlinks,
    citecolor=blue,
    filecolor=green,
    linkcolor=purple,
    urlcolor=red,
}

\usepackage{slashed}

\newcommand{\ie}{\begin{equation}}
\newcommand{\fe}{\end{equation}}
\newcommand{\se}{\begin{eqnarray}}
\newcommand{\ff}{\end{eqnarray}}

\usepackage{hyperref}
\hypersetup{colorlinks,breaklinks,
			citecolor=[rgb]{0,0.0,1.0},
            urlcolor=[rgb]{0.0,0.0,1.0},
            linkcolor=[rgb]{0,0.5,0.9}}

\begin{document}

\title{An exact stationary axisymmetric vacuum solution within a metric-affine bumblebee gravity}

\author{A. A. Ara\'{u}jo Filho}
\email{dilto@fisica.ufc.br}

\author{J. R. Nascimento}
\email{jroberto@fisica.ufpb.br}
\affiliation{Departamento de Física, Universidade Federal da Paraíba, Caixa Postal 5008, 58051-970, João Pessoa, Paraíba,  Brazil.}

\author{A. Yu. Petrov}
\email{petrov@fisica.ufpb.br}
\affiliation{Departamento de Física, Universidade Federal da Paraíba, Caixa Postal 5008, 58051-970, João Pessoa, Paraíba,  Brazil.}

\author{P. J. Porfírio}
\email{pporfirio@fisica.ufpb.br}
\affiliation{Departamento de Física, Universidade Federal da Paraíba, Caixa Postal 5008, 58051-970, João Pessoa, Paraíba,  Brazil.}




\date{\today}

\begin{abstract}

 Within the framework of the spontaneous Lorentz symmetry breaking (LSB), we consider a metric-affine generalization of the gravitational sector of the Standard Model Extension (SME), including the Lorentz-violating (LV) coefficients $u$ and $s^{\mu\nu}$. In this model, we derive the modified Einstein field equations in order to obtain a new axisymmetric vacuum spinning solution for a particular bumblebee's profile. Such a solution has the remarkable property of incorporating the effects of LSB through the LV dimensionless parameter $X=\xi b^2$, with $\xi$ is the nonminimal coupling constant, and $b^2=b^{\mu}b_{\mu}$, with $b_{\mu}$ is the vacuum expectation value of the bumblebee field; as the LSB is turned off, $X=0$, we recover the well-established result,  the Kerr solution, as expected. Afterwards, we calculate the geodesics, the radial acceleration and thermodynamic quantities for this new metric. We also estimate an upper bound for $X$ by using astrophysical data of the advance of Mercury's perihelion.

\end{abstract}


\maketitle


\section{Introduction}
\label{sec:intro}

The well-established Lorentz symmetry, arising from the principles of special relativity, implies that physical laws remain equivalent for all observers, given the condition of maintaining inertial frames. Embodying both rotational and boost symmetries, Lorentz invariance emerges as a foundational aspect, particularly relevant in the contexts of general relativity (GR) and the standard model of particle physics (SM). In particular, in curved spacetimes, the local Lorentz symmetry holds due to the Lorentzian character of the background. Conversely, the Lorentz invariance violation generally leads to the introduction of directional or velocity dependencies of physical variables, thereby inducing modifications in the dynamics of particles and waves \cite{STR1,STR2,STR3,STR4,STR5,STR6,STR7}.

Symmetry breaking processes present intriguing consequences that can serve as potential indicators of novel physical phenomena. Notably, LSB gives rise to a spectrum of distinctive features \cite{liberati2013,tasson2014,hees2016}, offering insights into the realm of quantum gravity \cite{rovelli2004}. Theoretical models, ranging from closed-string theories \cite{New1,New2,New3,New4,New5} and loop quantum gravity \cite{New6,New7} to noncommutative spacetimes \cite{New8,New9}, non-local gravity models \cite{Modesto:2011kw, Nascimento:2021bzb}, spacetime foam models \cite{New10,New11}, and (chiral) field theories defined on spacetimes with nontrivial topologies \cite{New12,New13,New14,New15}, as well as Ho\v{r}ava-Lifshitz gravity \cite{New16} and cosmology \cite{sv1,sv2}, often are based on the assumption of a departure from Lorentz invariance.

Investigations involving thermal aspects in the context of LSB could supply further information about the primordial Universe. In other words, this corroborates the fact that the size of the Universe at that stage was comparable with the characteristic scales of LSB \cite{kostelecky2011data}.
The thermal properties within the context of LSB have been initially proposed in \cite{colladay2004statistical}. After that, recently many works have been made in various scenarios, such as linearized gravity \cite{aa2021lorentz}, Pospelov and Myers-Pospelov \cite{araujo2021thermodynamic,anacleto2018lorentz} electrodynamics, CPT-even and CPT-odd LV terms \cite{casana2008lorentz,casana2009finite,araujo2021higher,aguirre2021lorentz}, higher-dimensional operators \cite{Mariz:2011ed, reis2021thermal}, bouncing universe \cite{petrov2021bouncing2}, \textit{rainbow} gravity \cite{furtado2023thermal}, and Einstein-aether theory \cite{aaa2021thermodynamics}.

Furthermore, the consistent implementation of LSB within the gravitational framework is extremely complicated in comparison with introducing LV extensions in non-gravitational field theories. In flat spacetimes, additive LV terms, such as the Carroll-Field-Jackiw \cite{CFJ} and aether terms \cite{aether, Gomes:2009ch} can be introduced, see \cite{colladay1998lorentz} for a generic approach incorporating all possible LV minimal couplings. Conversely, the application of such features in curved spacetimes encounters inherent complexities and requires special attention. 

Certainly, constant tensors are well defined in Minkowski spacetime; however, extending straightforward conditions for the tensors to be constant, like $\partial_{\mu}k_{\nu}=0$, to curved spacetimes proves to be a complicated task. The simplest condition $\partial_{\mu}k_{\nu}=0$ is clearly inconsistent with the general covariance requirement, while its natural covariant extension, $\nabla_{\mu}k_{\nu}=0$, imposes stringent constraints on spacetime geometries, known as the challenging no-go constraints \cite{kostelecky2021backgrounds}, which are notoriously difficult to satisfy. Consequently, the most appropriate manner of incorporating (local) LSB into gravitational theories involves the mechanism of spontaneous symmetry breaking. In this scenario, Lorentz/CPT violating coefficients (operators) emerge as vacuum expectation values of some dynamic tensor fields, influenced by nontrivial potentials.

The Standard Model Extension (SME), including its gravitational sector, is a general framework, proposed by Kosteleck{\`y} \cite{5}, which encompasses all conceivable coefficients for Lorentz/CPT violation. Specifically, within its gravitational sector, the SME 
is defined on a Riemann-Cartan manifold, wherein torsion is treated as a dynamic geometrical quantity alongside the metric. Despite the possibility of introducing non-Riemannian terms in the gravity SME sector, existing studies have predominantly focused on the metric approach to gravity, wherein the metric serves as the sole dynamical geometric field.

Within this picture, research efforts have primarily concentrated on deriving exact solutions for different models accommodating LSB in curved spacetimes. Examples include investigations into bumblebee gravity in the metric approach \cite{6,7,8,9,10,11,12,13,14,Maluf:2021lwh, KumarJha:2020ivj}, the Einstein-aether model \cite{15}, parity-violating models \cite{16,17,18,19,20, Rao:2023doc}, and Chern-Simons modified gravity \cite{21,22,23}. Experimental tests to detect LSB signals in the weak field regime of the gravitational field have also been carried out, with Solar System experiments being particularly noteworthy in this regard \cite{24,25,26}. However, the recent observation of gravitational waves in the LIGO/VIRGO collaboration \cite{LIGOScientific:2016aoc}, thanks to our current pace of technological growth, opened up a new window to probe the strong field regime of gravity which lays out a powerful tool to deeply understand the complex properties of compact objects, like black holes (pictures of their shadows have already been taken \cite{EventHorizonTelescope:2019dse, EventHorizonTelescope:2022wkp}). Moreover, it allows us to bring out a very thorny topic of checking GR at astrophysical scales. Rotating black holes play a pivotal role in whole this issue since the realistic (astrophysical) ones possess non-trivial angular momenta. In particular, finding new rotating black hole solutions in alternative theories of gravity remains a promising route to gather information on physics beyond GR \cite{rot1, rot2, rot3, rot4}. This is indeed our primary task in this present work.

While numerous works consider modified theories of gravity within the usual metric approach, there is growing interest in exploring more generic geometrical frameworks. Notably, in this environment, there are specific motivations for investigating theories of gravity within a Riemann-Cartan background, such as the induction of gravitational topological terms (see f.e. \cite{Nascimento:2021vou}). Another intriguing non-Riemannian geometry is the Finsler one \cite{Bao}, which has been extensively linked to LSB in a variety of studies \cite{Foster, KosE, Sch1, CollM, Sch2}.

The metric-affine (Palatini) formalism stands out as the most compelling generalization of the metric approach, in which the metric and connection are regarded as independent dynamical geometrical quantities (for a comprehensive discussion and intriguing findings within the Palatini approach, see e.g., \cite{Ghil1,Ghil2}, and references therein). In spite of the advancements in this framework, LSB remains relatively unexplored in this context. However, recent works have begun to fill this gap, particularly within the context of bumblebee gravity scenarios \cite{Paulo2, Paulo3, Paulo4}. Remarkably, the authors have derived the field equations, generically solved them, and explored stability conditions and associated dispersion relations for various matter sources in the weak field and post-Newtonian limit. Additionally, at the quantum level, they have computed the divergent piece of one-loop corrections to the spinor effective action through two distinct methodologies: utilizing the diagrammatic method in the weak gravity regime and, more generally, employing the Barvinsky-Vilkovisky technique. In particular, an exact Schwarzschild-like solution has been found in \cite{Filho:2022yrk} and estimations for the LV parameter have been provided from classical gravitational tests. Furthermore, the shadow and the quasinormal modes of this black hole have also been obtained in \cite{Lambiase:2023zeo, Jha:2023vhn,hassanabadi2023gravitational}.

Similarly, a metric-affine version of Chern-Simons modified gravity, invariant under projective transformations, has been proposed \cite{Paulo5, Boudet1, Boudet2}. In this context, the authors have adopted a perturbative scheme to solve the field equations, given the elusive nature of an exact solution to the connection equation. Furthermore, analyses of quasinormal modes of Schwarzschild black holes have been conducted within this model, further enriching the exploration of gravitational phenomena.

Here, we focus on a particular metric-affine bumblebee gravity. This model can be connected with the LV coefficients of the SME by assuming $u$ and $s^{\mu\nu}$ to be non-trivial, while $t^{\mu\nu\alpha\beta}=0$. In this work, we obtain a stationary and axisymmetric vacuum rotating solution, which is the first one found in this context. Afterwards, we perform the thermodynamic calculations, the geodesics, the radial acceleration, and the estimations for LV coefficients by using experimental data from the advance of Mercury's perihelion.

The structure of the paper is organized as follows. In Sec. \ref{traceless}, we define the metric-affine bumblebee gravity model under consideration and derive its respective equations of motion. In Sec. \ref{Application}, we obtain a stationary axisymmetric solution, representing itself a generalization of the Kerr metric and some applications, concerning the thermodynamic state quantities, radial acceleration, the geodesics, and the estimations for the LSB parameter by using experimental data from the advance of Mercury's perihelion are also provided. Finally, in Sec. \ref{summary}, we present our conclusions.


\section{The traceless metric-affine bumblebee gravity model}
\label{traceless}

  We here briefly review the traceless metric-affine bumblebee gravity model earlier discussed in \cite{Paulo2, Paulo3, Paulo4}. To begin with, the action of this model reads
    \begin{eqnarray}
      \mathcal{S}_{B}&=&\int d^{4}x\,\sqrt{-g}\left[\frac{1}{2\kappa^2}\left(R(\Gamma)+\xi\left(B^{\mu}B^{\nu}-\frac{1}{4}B^{2}g^{\mu\nu}\right)R_{\mu\nu}(\Gamma)\right)-\frac{1}{4}B^{\mu\nu}B_{\mu\nu}-\right.\nonumber\\ &-&\left.V(B^{\mu}B_{\mu}\pm b^{2})\right]+
      \int d^{4}x\sqrt{-g}\mathcal{L}_{mat}(g_{\mu\nu},\psi),
      \label{bumb}
  \end{eqnarray}
    where the geometrical quantities $R(\Gamma) \equiv g_{\mu\nu} R^{\mu\nu}(\Gamma), R^{\mu\nu}(\Gamma)$ and $R^{\mu}_{\,\,\,\nu\alpha\beta}(\Gamma)$ are the Ricci
scalar, Ricci tensor and Riemann tensor. Note that the above action is defined in the metric-affine (Palatini) approach, where the connection is assumed to be an independent entity of the metric. The matter Lagrangian $\mathcal{L}_{mat}(g_{\mu\nu}, \psi)$, where $\psi$ stands for matter fields, is supposed to involve coupling of matter with the metric only. The vector field $B_{\mu}$ is the bumblebee one,  $B_{\mu\nu}=(\mathrm{d}B)_{\mu\nu}$ is the field strength and $B^{2}\equiv g^{\mu\nu}B_{\mu}B_{\nu}$. Another striking feature of the bumblebee model is the presence of the potential  $V(B^{\mu}B_{\mu}\pm b^2)$ with a non-trivial vacuum expectation value (VEV), $<B_{\mu}>=b_{\mu}$, where $b_{\mu}$ represents a particular minimum of this potential. Such a mechanism allows for the spontaneous LSB. The constant $b^2$ is defined as $b^2\equiv g^{\mu\nu}b_{\mu}b_{\nu}$. Drawing a parallel with the SME \cite{kostelecky2004gravity}, our model can be cast into a compact form, namely:
\begin{equation}
 \begin{split}
     \mathcal{S}_{B}&=\int d^{4}x\,\sqrt{-g}\bigg\{\frac{1}{2\kappa^2}\left[\left(1-u\right)R(\Gamma)+s^{\mu\nu}R_{\mu\nu}(\Gamma)+t^{\mu\nu\alpha\beta}R_{\mu\nu\alpha\beta}\right]-\frac{1}{4}B^{\mu\nu}B_{\mu\nu}-\\
     &-V(B^{\mu}B_{\mu}\pm b^{2})\bigg\}+\int d^{4}x\sqrt{-g}\mathcal{L}_{mat}(g_{\mu\nu},\psi),
     \end{split}
     \label{S2}
 \end{equation}
where $u$, $s^{\mu\nu}$ and $t^{\mu\nu\alpha\beta}$ are LV coefficients. Doing a straightforward comparison with our model, we conclude that
\begin{eqnarray}
    u=0, \,\,\, s^{\mu\nu}=\xi\left(B^{\mu}B^{\nu}-\frac{1}{4}B^2 g^{\mu\nu}\right) \,\,\, \mbox{and}\,\,\, t^{\mu\nu\alpha\beta}=0
\end{eqnarray}
or, equivalently,
\begin{eqnarray}
    u=\frac{\xi}{4}B^2,\,\,\, s^{\mu\nu}=\xi B^{\mu}B^{\nu}\,\,\, \mbox{and}\,\,\, t^{\mu\nu\alpha\beta}=0.
    \label{sec}
\end{eqnarray}
  The difference between both representations is that the traceless piece of $s^{\mu\nu}$ has been absorbed into the definition of $u$ in Eq.(\ref{sec}).

It is worth stressing out that the above action is invariant under projective transformations of the connection,
\begin{equation}
\Gamma^{\mu}_{\nu\alpha}\longrightarrow \Gamma^{\mu}_{\nu\alpha}+\delta^{\mu}_{\alpha}A_{\nu},
\label{Proj}
\end{equation}
where $A_{\alpha}$ is an arbitrary vector. It is easy to check that the Riemann tensor under the projective transformation given by (\ref{Proj}), changes as follows:
\begin{equation}
    R^{\mu}_{\,\,\,\nu\alpha\beta}\longrightarrow R^{\mu}_{\,\,\,\nu\alpha\beta}-2\delta^{\mu}_{\nu}\partial_{[\alpha}A_{\beta]},
\end{equation}
as a consequence, the symmetric part of the Ricci tensor is invariant under Eq.(\ref{Proj}), as well as, the whole action (\ref{S2}).

The model given by the action (\ref{S2}) belongs to a more generic class of gravitational theories called Ricci-based ones \cite{Afonso:2017bxr, BeltranJimenez:2017doy, Delhom:2021bvq}. It has been shown that, for this class of models, the projective invariance avoids the emergence of gravitational ghost-like propagating degrees of freedom \cite{AD}.

  \subsection{Field equations}
\label{FEI}


\subsubsection{The connection equation}



By varying the action \eqref{bumb} with respect to the connection, we obtain
\begin{equation}
\nabla^{(\Gamma)}_{\alpha}  \left[ \sqrt{-h} h^{\mu\nu}  \right] = \sqrt{-h}\left[T^{\mu}_{\,\,\alpha\lambda}h^{\nu\lambda}+T^{\lambda}_{\,\,\lambda\alpha}h^{\mu\nu}-\frac{1}{3}T^{\lambda}_{\,\,\lambda\beta}h^{\nu\beta}\delta^{\mu}_{\alpha}\right], \label{fieldequation}
\end{equation}
where $T^{\mu}_{\,\,\alpha\lambda}=\Gamma^{\mu}_{\,\,\alpha\lambda}-\Gamma^{\mu}_{\,\,\lambda\alpha}$ is the torsion tensor. Furthermore, we defined
\begin{equation}
    h^{\mu\nu}=\frac{1}{\sqrt{\mathrm{\det\hat{\Omega}^{-1}}}}g^{\mu\alpha}\left(\Omega^{-1}\right)^{\nu}_{\alpha},
\end{equation}
where we have defined the inverse of the deformation matrix by $\hat{\Omega}^{-1}\equiv \left(1-\frac{\xi}{4}B^2\right)\hat{I}+\xi\hat{BB}$ and  
$\left(\Omega^{-1}\right)^{\nu}_{\alpha}\equiv\left(1-\frac{\xi}{4}B^2\right)\delta^{\nu}_{\alpha}+\xi B^{\nu}B_{\alpha}$. As discussed in \cite{BeltranJimenez:2019acz},  the torsion tensors on the {\it r.h.s} of Eq.(\ref{fieldequation}) are pure gauge ones, i.e., they can be eliminated through an appropriate gauge-fixing choice \cite{BeltranJimenez:2017doy}. Thus, disregarding the gauge modes, the solution of the connection equation is simply given by the Levi-Civita connection of the $h$-metric, namely,
\begin{equation}
\Gamma\indices{^\mu_\nu_\alpha} = \left\{ \indices{^\mu_\nu_\alpha} \right\}^{(h)} = \frac{1}{2}h^{\mu\lambda}\left(-\partial_{\lambda}h_{\nu\alpha}+\partial_{\nu}h_{\alpha\lambda}+\partial_{\alpha}h_{\lambda\nu}\right),
\label{conn}
\end{equation}
where $h^{\mu\nu}$ is the inverse metric of $h_{\mu\nu}$.
 
Through a direct calculation, one obtains that
\begin{equation}
    \mathrm{\det\hat{\Omega}^{-1}}=\left(1-\frac{\xi B^2}{4}\right)^{3}\left(1+\frac{3}{4}\xi B^2\right).
\end{equation}
Using the previous result, we can find
\begin{eqnarray}
    h^{\mu\nu}=\frac{1}{\sqrt{\left(1-\frac{\xi B^2}{4}\right)\left(1+\frac{3}{4}\xi B^2\right)}}\left[g^{\mu\nu}+\frac{\xi B^{\mu}B^{\nu}}{\left(1-\frac{\xi B^2}{4}\right)}\right].
    \label{metric1}
\end{eqnarray}
Similarly, we have
\begin{equation}
    h_{\mu\nu}=\sqrt{\left(1+\frac{3}{4}\xi B^2\right)\left(1-\frac{\xi B^2}{4}\right)}g_{\mu\nu}-\xi\sqrt{\frac{\left(1-\frac{\xi B^2}{4}\right)}{\left(1+\frac{3}{4}\xi B^2\right)}} B_{\mu}B_{\nu}.
    \label{metric2}
\end{equation}

\subsubsection{The metric equation}

The metric equation is obtained by varying the action \eqref{bumb} with respect to $g_{\mu\nu}$. By doing so, we get 
\begin{equation}
\begin{split}
    &\left(1-\frac{\xi B^2}{4}\right)R_{(\mu\nu)}(\Gamma)-\frac{1}{2}g_{\mu\nu}R(\Gamma)+2\xi\left[B^{\alpha}B_{(\mu}R_{\nu)\alpha}(\Gamma)\right]-\frac{\xi}{4}B_{\mu}B_{\nu}R(\Gamma)-\\
    &-\frac{\xi}{2}g_{\mu\nu}B^{\alpha}B^{\beta}R_{\alpha\beta}(\Gamma)+
    \frac{\xi}{8}B^{2}g_{\mu\nu}R(\Gamma)=\kappa^2 T_{\mu\nu},
    \end{split}
    \label{kjh}
    \end{equation}
where the stress-energy tensor $T_{\mu\nu}=T_{\mu\nu}^{mat}+T_{\mu\nu}^{B}$, with
\begin{eqnarray}
    T_{\mu\nu}^{mat}=-\frac{2}{\sqrt{-g}}\frac{\delta (\sqrt{-g}\mathcal{L}_{mat})}{\delta g^{\mu\nu}}
\end{eqnarray}
 and 
\begin{equation}
    T_{\mu\nu}^{B}=B_{\mu\sigma}B_{\nu}^{\,\,\sigma}-\frac{1}{4}g_{\mu\nu}B^{\alpha}_{\,\,\sigma}B^{\sigma}_{\,\,\alpha}-Vg_{\mu\nu}+2V^{\prime}B_{\mu}B_{\nu}.
    \end{equation}

 By contracting Eq.\eqref{kjh} with $g^{\mu\nu}$, $B^{\mu}$ and $B^{\mu}B^{\mu\nu}$, we are able to find important relations, namely,
\begin{eqnarray}
R(\Gamma)&=&-\kappa^2 T;\label{RT}\\
    \nonumber B^{\mu} R_{\mu\nu}(\Gamma) &=& \frac{4 \kappa^{2}}{(4+3\xi B^{2})}\bigg\{ T_{\mu\nu} B^{\mu}-\frac{B_{\nu}T}{2} - \frac{2\xi B_{\nu}}{4 + 5\xi B^{2}}\bigg[ B^{\alpha}B^{\beta}T_{\alpha\beta} - \\
    &-&\frac{1}{4}B^{2} T \left( 1 - \frac{3}{4}\xi B^{2} \right)    \bigg]      \bigg\};\label{b1}\\
    B^{\mu}B^{\nu} R_{\mu\nu}(\Gamma) &=& \frac{4\kappa^{2}}{4+5\xi B^{2}} \left[ B^{\mu}B^{\nu}T_{\mu\nu} -\frac{B^{2}T}{8} (4 + \xi B^{2})        \right]. \label{b2}
\end{eqnarray}
Now substituting  Eqs.(\ref{RT}), (\ref{b1}), (\ref{b2}) into Eq.(\ref{kjh}), this results in
\begin{equation}
\begin{split}
    R_{\mu\nu}(h)&=\kappa^{2}_{eff}\bigg\{T_{\mu\nu}-\frac{1}{2}g_{\mu\nu}T+\frac{2\xi g_{\mu\nu}}{(4+5\xi B^{2})} \left[ B^{\alpha}B^{\beta}T_{\alpha\beta} -\frac{B^{2}T}{16} (4 - 3\xi B^{2})        \right]+\\
    &+\frac{8\xi}{(4+3\xi B^2)}B_{(\mu}\left[T_{\nu)\alpha}B^{\alpha}-\frac{B_{\nu)}T}{2}-\right.\\
    &-\left.\frac{2\xi B_{\nu)}}{(4+5\xi B^2)}\left(B^{\alpha}B^{\beta}T_{\alpha\beta}-\frac{1}{4}B^{2}T\left(1-\frac{3}{4}\xi B^2\right)\right)\right]\bigg\},
    \label{RR}
    \end{split}
\end{equation}
where $\kappa^{2}_{eff}=\dfrac{\kappa^{2}}{1-\frac{\xi B^2}{4}}$.

Notice that the \textit{r.h.s.} of Eq.\eqref{RR} can be completely written in terms of the $h$-metric by using  Eqs.(\ref{metric1}) and (\ref{metric2}), rendering it a dynamical equation for $h_{\mu\nu}$. However, for our purposes, its explicit form is irrelevant.

\subsubsection{The bumblebee field equation}

Let us now turn our attention to the bumblebee field equation. Varying the action (\ref{bumb}) with respect to $B_{\mu}$, one finds
\begin{equation}
    \nabla^{(g)}_{\mu}B^{\mu\alpha}=-\frac{\xi}{\kappa^2}g^{\nu\alpha}B^{\mu}R_{\mu\nu}(\Gamma)+\frac{\xi}{4\kappa^2}B^{\alpha}R(\Gamma)+2V^{\prime}(B^{\mu}B_{\mu}\pm b^2)B^{\alpha},
    \label{bumb2}
\end{equation}
where the prime above stands for the derivative with respect to the argument of the potential $V$, and $\nabla^{(g)}_{\mu}$ is the covariant derivative defined in terms of the Levi-Civita connection of $g_{\mu\nu}$. Inserting Eqs.(\ref{RT}) and (\ref{b1}) into Eq. (\ref{bumb2}), one gets a Proca-like equation
\begin{eqnarray}
\nabla^{(g)}_{\mu}B^{\mu\alpha}=\mathcal{M}^{\alpha}_{\,\,\,\nu}B^{\nu},
\label{proca}
\end{eqnarray}
where we have defined the effective mass-squared tensor by
\begin{eqnarray}
\nonumber \mathcal{M}^{\alpha}_{\,\,\,\nu}&=&\bigg\{2V^{\prime}+\frac{\xi T\left(4-3\xi B^2\right)}{4\left(4+3\xi B^2\right)}+\frac{8\xi^2}{\left(4+3\xi B^2\right)\left(4+5\xi B^2\right)}\bigg[B^{\mu}B^{\lambda}T_{\mu\lambda}-\\
&-&\frac{1}{4}B^2 T\left(1-\frac{3}{4}\xi B^2 \right)\bigg]\bigg\}\delta^{\alpha}_{\,\,\,\nu}-\frac{4\xi}{\left(4+3\xi B^2\right)}T^{\alpha}_{\,\,\,\nu}.
\label{Mat}
\end{eqnarray}
Note that the new unconventional interaction terms between the bumblebee field and the stress-energy tensor allow us to have new effects, essentially different from the metric case. One can mention, for example, the mechanism of spontaneous vectorization that occurs when the bumblebee field spontaneously acquires an effective mass near high-density compact objects \cite{Ramazanoglu:2017xbl, Ramazanoglu:2019jrr, Cardoso:2020cwo}. Moreover, due to the negative sign between the first and second terms in Eq.(\ref{Mat}), the determinant of the effective mass-squared matrix can assume negative values leading to tachyonic-like instabilities. 

 Observe, however, that Eq. (\ref{proca}) can be cast into a more convenient form by introducing a conserved current, $J^{\mu}$. To see that in more detail, let us take the divergence of Eq.(\ref{proca}), and then one obtains
\begin{equation}
    \nabla^{(g)}_{\mu}J^{\mu}=0,
    \label{fg}
\end{equation}
where 
\begin{equation}
    J^{\mu}=\mathcal{M}^{\mu}_{\,\,\,\nu}B^{\nu}.
\end{equation}
By defining the bumblebee field equation in terms of a conservation of a current, Eq.(\ref{fg}), it permits us to find regular solutions more easily. We shall discuss the exact solutions of the metric-affine bumblebee model in the next section.

\section{Applications: a stationary axisymmetric solution in metric-affine traceless bumblebee model}
\label{Application}
By defining the bumblebee field equation in terms of the conservation of a current, Eq.(\ref{fg}), it permits us to find regular solutions more easily. We shall discuss on exact solutions of the metric-affine bumblebee model below. In particular, we are interested in stationary axisymmetric solutions for the metric-affine traceless bumblebee model discussed before. Initially, let us restrict our attention to vacuum solutions which are featured by the absence of matter sources, $T_{\mu\nu}^{(mat)}=0$. Apart from that, we fix the bumblebee field to assume its vacuum expectation value, i.e., $<B_{\mu}>=b_{\mu}$, that leads to $V=0$ and $V^{\prime}=0$.

In this scenario, we shall start with the field equations displayed in the last subsection. The first important ingredient is the metric. Notice that Eq.(\ref{RR}) is the dynamical equation for the metric  $h_{\mu\nu}$. Therefore, it is more convenient to manipulate the field equations in the Einstein frame. We focus on a particular sort of stationary axisymmetric metric, the well-known Kerr one, its line element in Boyer-Lindquist coordinates $(t,r,\theta,\phi)$ is given by
\begin{eqnarray}
    \nonumber \mathrm{d}s^2_{(h)}&=&-\left(\frac{\Delta-a^2\sin^2{\theta}}{\rho^2}\right)\mathrm{d}t^2-\frac{4aMr\sin^2{\theta}}{\rho^2}\mathrm{d}t\mathrm{d}\phi+\frac{\rho^2}{\Delta}\mathrm{d}r^2+\rho^2 \mathrm{d}\theta^2+\\
    &+& \left(\frac{(r^2 +a^2)^2-a^2 \Delta\sin^2{\theta}}{\rho^2}\right)\sin^2 \theta \mathrm{d}\phi^2,
    \label{Kerr}
\end{eqnarray}
where $\Delta=r^2 +a^2 -2Mr$ and $\rho^2=r^2+a^2 \cos^2 \theta$. The second ingredient is the form of $b_{\mu}$. In order to find a regular solution, we impose that the norm of the conserved current in the Einstein frame, $J^{2}=h^{\mu\nu}J_{\mu}J_{\nu}$, vanishes throughout the spacetime, which guarantees that the current does not diverge at the horizon \footnote{A similar choice has been done in the context of Galileons in \cite{Rinaldi, Babichev}.}. Such a requirement is fulfilled assuming $b_{\mu}$ to have the form:
\begin{equation}
    b_{\mu}=[0,b(r),c(\theta),0],
    \label{bumbp}
\end{equation}
 which leads to the vanishing of the field strength associated with it, $b_{\mu\nu}=\left(db\right)_{\mu\nu}$. As a consequence, $T_{\mu\nu}$ and also $J_{\mu}$ vanish even without imposing any previous condition on $b(r)$ and $c(\theta)$. In this scenario, the field equations \eqref{RR} reduce to
 \begin{equation}
     R_{\mu\nu}(h)=0,
 \end{equation}
 whose stationary axisymmetric solution is given by the Kerr metric \eqref{Kerr}. Before proceeding further, it is worth calling attention to the conventions that we shall adopt here: tilded objects are defined in the Einstein frame, namely, their indices can be raised or lowered using the auxiliary metric, $h^{\mu\nu}$. For example, $\tilde{b}^{\mu}=h^{\mu\nu}b_{\nu}$. Note that although $b^2=g^{\mu\nu}b_{\mu}b_{\nu}$ possesses an explicit dependence of $g^{\mu\nu}$, one can define a new object $\tilde{b}^2=h^{\mu\nu}b_{\mu}b_{\nu}$, which depends on $h_{\mu\nu}$. They are algebraically related to each other by $\tilde{b}^{2}=b^2\dfrac{\left(1+\frac{3 X}{4}\right)^{1/2}}{\left(1-\frac{X}{4}\right)^{3/2}}$, where $X \equiv \xi b^2$. Thereby, $b^2$ can properly be written in terms of $\tilde{b}^2$.  Furthermore, as we mentioned before, $b^2$ is a real constant, as $\tilde{b}^{2}$ is.

  The requirement that $\tilde{b}^{2}=const$ leads to $b(r)=\dfrac{|\tilde{b}|}{\sqrt{1-\frac{2M}{r}+\frac{a^2}{r^2}}}$ and $c(\theta)=|\tilde{b}|a\cos{\theta}$, thus the VEV is given by
  \begin{equation}
      b_{\mu}=\left[0,\dfrac{|\tilde{b}|}{\sqrt{1-\frac{2M}{r}+\frac{a^2}{r^2}}},|\tilde{b}|a\cos{\theta},0\right].
  \end{equation}
 Notice that although $b_{\mu}$ diverges at the horizon -- due to an effect of a ``bad'' coordinates choice -- the physical observables are characterized by the scalar invariants built up from $b_{\mu}$ which are finite at the horizon. For example, $b^2=const$, and $J^2=0$, by construction, and $b^{\mu}b^{\nu}R_{\mu\nu}=0$.

In order to find the metric $g_{\mu\nu}$, we substitute Eq.\eqref{Kerr} in Eq.(\ref{metric2}), identifying $B_{\mu}=b_{\mu}$. After that, one obtains the line element for $g_{\mu\nu}$, namely,
\begin{eqnarray}
    \nonumber \mathrm{d}s^2_{(g)}&=&-\left(\frac{\Delta-a^2\sin^2{\theta}}{\rho^2}\right)\frac{\mathrm{d}t^2}{\sqrt{\left(1+\frac{3X}{4}\right)\left(1-\frac{X}{4}\right)}}-\frac{4aMr\sin^2{\theta}}{\sqrt{\left(1+\frac{3X}{4}\right)\left(1-\frac{X}{4}\right)}\rho^2}\mathrm{d}t \mathrm{d}\phi+\\  
   \nonumber &+&\frac{1}{\Delta \sqrt{\left(1+\frac{3X}{4}\right)\left(1-\frac{X}{4}\right)}}\left(a^2 \cos^2 \theta+r^2 \dfrac{\left(1+\frac{3 X}{4}\right)}{\left(1-\frac{X}{4}\right)}\right)\mathrm{d}r^2 +\\
   \nonumber &+&\frac{1}{\sqrt{\left(1+\frac{3X}{4}\right)\left(1-\frac{X}{4}\right)}}\left(r^2 + a^2 \cos^2 {\theta} \dfrac{\left(1+\frac{3 X}{4}\right)}{\left(1-\frac{X}{4}\right)} \right) \mathrm{d}\theta^2
    +\\
    \nonumber &+&\frac{(r^2 +a^2)^2-a^2 \Delta\sin^2{\theta}}{\sqrt{\left(1+\frac{3X}{4}\right)\left(1-\frac{X}{4}\right)}\rho^2}\sin^2 \theta \mathrm{d}\phi^2 +\\
    &+&\frac{2r X a\cos{\theta} }{\sqrt{\left(1+\frac{3X}{4}\right)}\left(1-\frac{X}{4}\right)^{\frac{3}{2}}}\frac{\mathrm{d}r\mathrm{d}\theta}{\sqrt{\Delta}},
    \label{metric3}
\end{eqnarray}
where the deviations from the standard Kerr solution are clear and manifest themselves as corrections depending on the LV coefficient $X$ arising within the metric-affine bumblebee gravity. It is worth pointing out that this is the primary result of this work. Note that the line element in Eq.(\ref{metric3}) can be viewed as a LV modified Kerr metric. The presence of the LV coefficient affects all components of the metric $g_{\mu\nu}$, as might be explicitly deduced from the previous line element.
As for the Kretchmann invariant, while its qualitative behavior is rather similar to that one in our previous paper \cite{Filho:2022yrk}, its explicit form is much more complicated. So, to gain further insight into this solution, let us treat two different asymptotic cases. First, in the far-field limit, the line element \eqref{metric3} reads
\begin{eqnarray}
    \nonumber \mathrm{d}s^2_{(g)}&=&-\left(\dfrac{1-\frac{2M}{r}}{\sqrt{\left(1+\frac{3X}{4}\right)\left(1-\frac{X}{4}\right)}}+\mathcal{O}\left(\frac{1}{r^2}\right)\right)\mathrm{d}t^2-\nonumber\\ 
    &-&\left(\frac{4aM\sin^2{\theta}}{r\sqrt{\left(1+\frac{3X}{4}\right)\left(1-\frac{X}{4}\right)}}+\mathcal{O}\left(\frac{1}{r^3}\right)\right)\mathrm{d}t \mathrm{d}\phi\nonumber\\
   \nonumber &+&\left(\frac{\sqrt{1+\frac{3X}{4}}}{\left(1-\frac{X}{4}\right)^{\frac{3}{2}}}+\mathcal{O}\left(\frac{1}{r}\right)\right)\mathrm{d}r^{2}+\left(\frac{r^2}{\sqrt{\left(1+\frac{3X}{4}\right)\left(1-\frac{X}{4}\right)}}+\mathcal{O}\left(\frac{1}{r}\right)\right)\left(\mathrm{d}\theta^2+\sin^{2}\theta \mathrm{d}\phi^2\right)\\
   &+&\frac{2Xa\cos\theta}{\sqrt{\left(1+\frac{3X}{4}\right)}\left(1-\frac{X}{4}\right)^{\frac{3}{2}}}\mathrm{d}r \mathrm{d}\theta.
\end{eqnarray}
It should be noted that this metric is not the standard asymptotic limit of the axially symmetric rotating metric due to the corrections depending on $X$. The second interesting limit consists of investigating the slow rotation regime of \eqref{metric3}, $\frac{a}{M}<<1$, in this case, the line element \eqref{metric3} becomes
\begin{eqnarray}
    \nonumber \mathrm{d}s^2_{(g)}&=&\left[-\dfrac{1-\frac{2M}{r}}{\sqrt{\left(1+\frac{3X}{4}\right)\left(1-\frac{X}{4}\right)}}-\frac{a^2}{r^3\sqrt{\left(1+\frac{3X}{4}\right)\left(1-\frac{X}{4}\right)}}\left(2M\cos^2 \theta\right)+\mathcal{O}(a^4)\right]\mathrm{d}t^2+\\
   \nonumber &+&\left[-\frac{4Ma\sin^{2}\theta}{r\sqrt{\left(1+\frac{3X}{4}\right)\left(1-\frac{X}{4}\right)}}+\mathcal{O}(a^3)\right]\mathrm{d}t\mathrm{d}\phi+\Bigg[\frac{1}{\left(1-\frac{2M}{r}\right)}\frac{\sqrt{1+\frac{3X}{4}}}{\left(1-\frac{X}{4}\right)^{\frac{3}{2}}}+\frac{a^2}{r^2 \left(1-\frac{2M}{r}\right)}\\
   \nonumber &\times &\Bigg(-\frac{1}{\left(1-\frac{2M}{r}\right)}\frac{\sqrt{1+\frac{3X}{4}}}{\left(1-\frac{X}{4}\right)^{\frac{3}{2}}}
   +\frac{\cos^2 \theta}{\sqrt{\left(1+\frac{3X}{4}\right)\left(1-\frac{X}{4}\right)}}\Bigg)+\mathcal{O}(a^4 )\Bigg]\mathrm{d}r^2+\\
   \nonumber&+&\left[\frac{r^2}{\sqrt{\left(1+\frac{3X}{4}\right)\left(1-\frac{X}{4}\right)}}+a^2 \cos^2 \theta \frac{\sqrt{1+\frac{3X}{4}}}{\left(1-\frac{X}{4}\right)^{\frac{3}{2}}}\right]\mathrm{d}\theta^2 +\\
   \nonumber&+&\left[\frac{r^2 \sin^2 \theta}{\sqrt{\left(1+\frac{3X}{4}\right)\left(1-\frac{X}{4}\right)}}+\frac{a^2 \sin^2 \theta}{\sqrt{\left(1+\frac{3X}{4}\right)\left(1-\frac{X}{4}\right)}}\left(1+\frac{2M}{r}\sin^2 \theta\right)+\mathcal{O}(a^4)\right]\mathrm{d}\phi^2 +\\
   &+&\left(\frac{2aX\cos\theta}{\sqrt{1-\frac{2M}{r}}\sqrt{\left(1+\frac{3X}{4}\right)}\left(1-\frac{X}{4}\right)^{\frac{3}{2}}}+\mathcal{O}(a^3)\right)\mathrm{d}r\mathrm{d}\theta.
\end{eqnarray}
Note that the zeroth-order piece in $a$ of the above line element recovers, after a suitable rescaling in the radial coordinate and the mass, the result found in \cite{Filho:2022yrk} for the spherically symmetric solution with LSB. Therefore, one concludes that the leading-order corrections to the spherically symmetric solution are boosted by linear and quadratic contributions in $a$.


\subsection{Thermodynamics}

\subsubsection{The Hawking temperature}

In order to accomplish the analysis of the event horizon, we consider the limit $1/g_{rr}\to 0$, which leads to
\ie
r_{\pm} = M \pm \sqrt{-a^2 + M^2},
\fe
when $\sin \theta =0$. Clearly, there is no modification associated with the LV parameters. It is worth mentioning that the same absence of the LV effects occurs for the angular velocity as well, i.e., $\Omega = -g_{t\phi}/g_{\phi\phi}$. In addition, to calculate the corresponding \textit{Hawking} temperature, let us take advantage of using the first law of thermodynamics, which reads
\ie
\mathrm{d}M = T \mathrm{d}S + \Omega\, \mathrm{d}J.
\fe
With it, the \textit{Hawking} temperature can straightforwardly be derived as 
\ie
T = \frac{\mathrm{d}M}{\mathrm{d}S}  = \frac{1}{2\pi r_{+}} \frac{\mathrm{d}M}{\mathrm{d}r_{+}}   = \frac{\sqrt{\frac{\left(a^2+r_{+}^2\right)^2}{4 r_{+}^2}-a^2}}{2 \pi  \left(a^2+r_{+}^2\right)},
\fe
where we have assumed that parameter $a$ is smaller than
$r_{+}$ and $M$. Notice that there is no change in the \textit{Hawking} temperature as well.

\subsubsection{The entropy}

\begin{figure}
    \centering
     \includegraphics[scale=0.4]{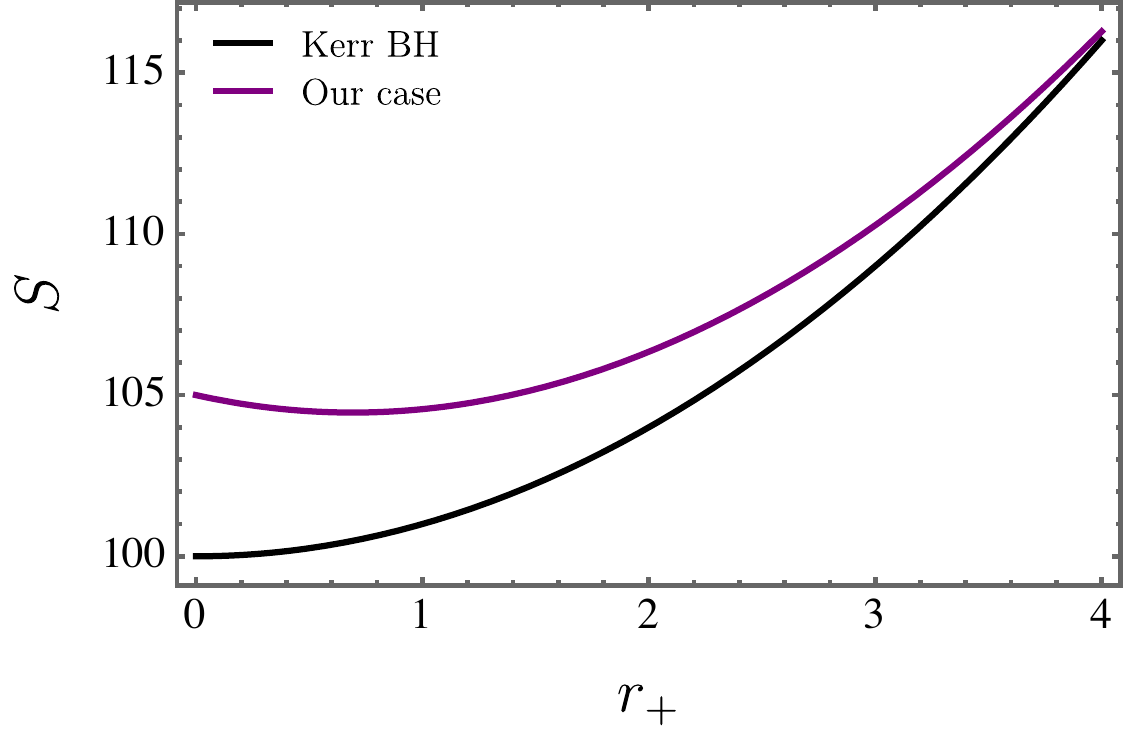}
    \includegraphics[scale=0.4]{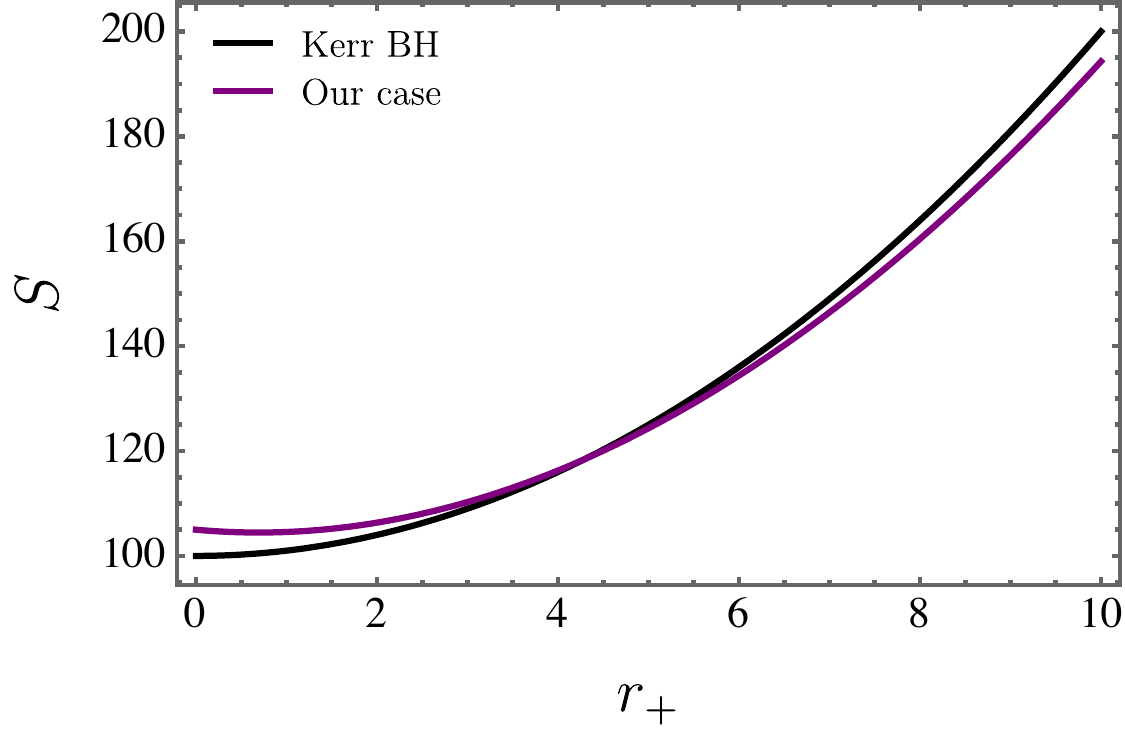}
    \caption{Comparison of the entropy for Kerr black hole with our case, varying the event horizon when $a=10$ and $X=0.2$.}
    \label{entropycomparison}
\end{figure}

\begin{figure}
    \centering
     \includegraphics[scale=0.4]{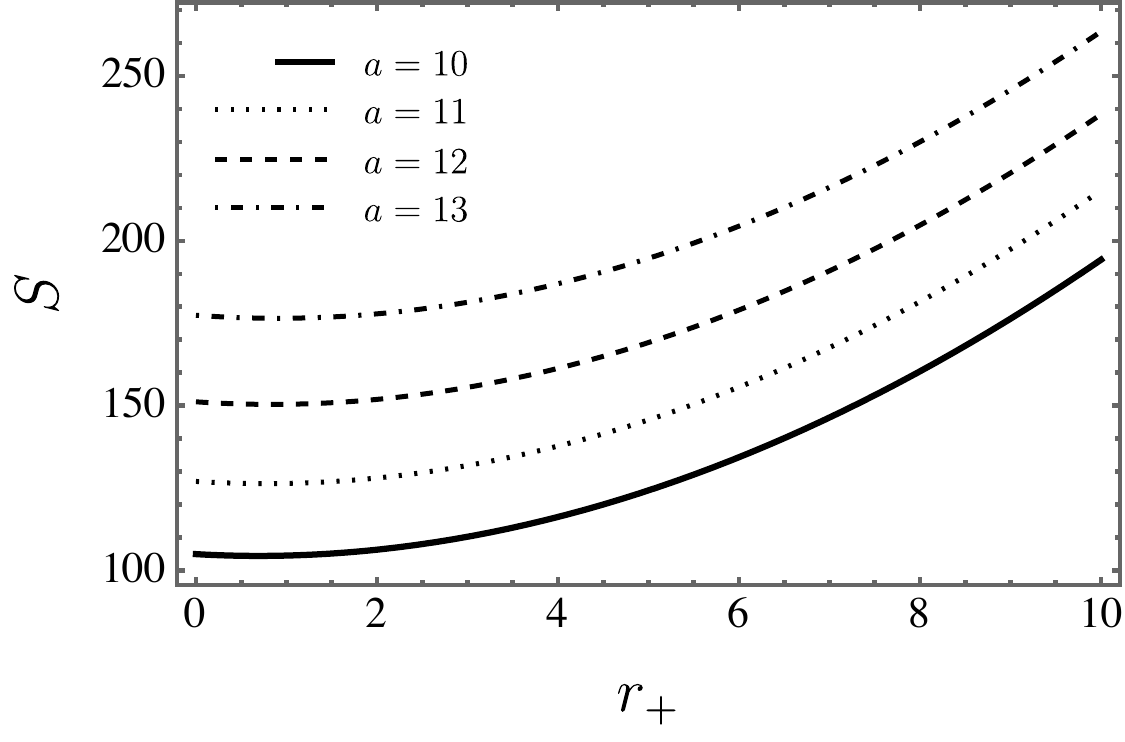}
     \includegraphics[scale=0.4]{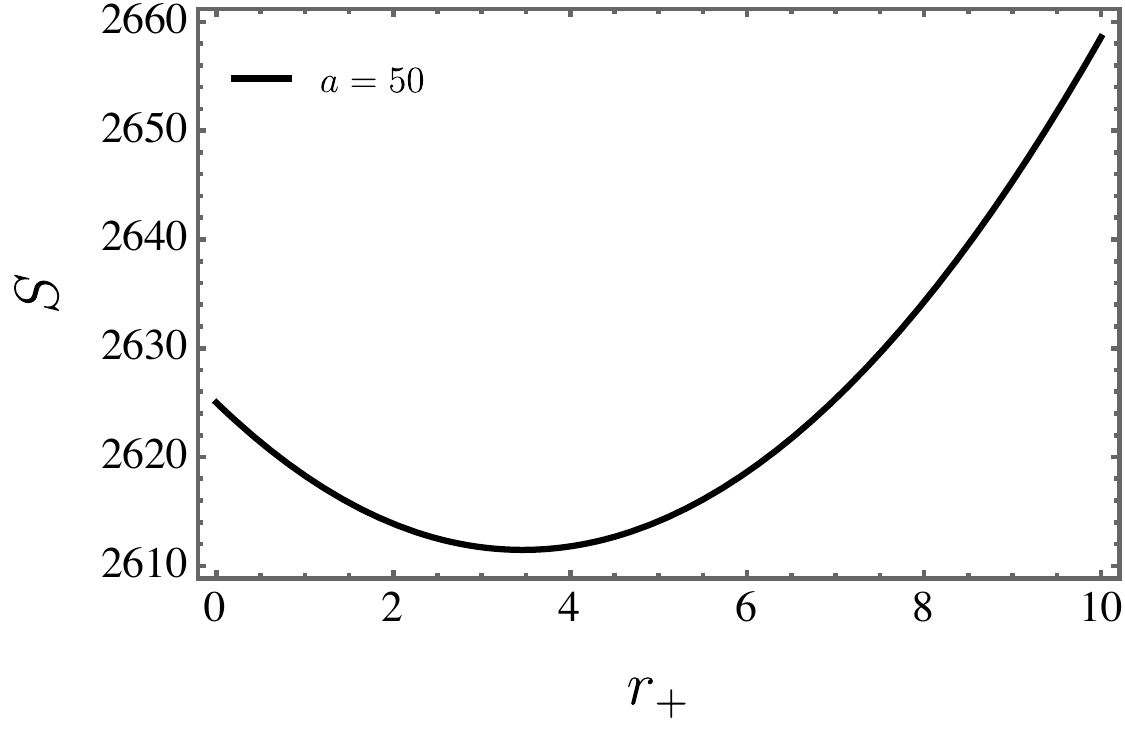}
     \includegraphics[scale=0.4]{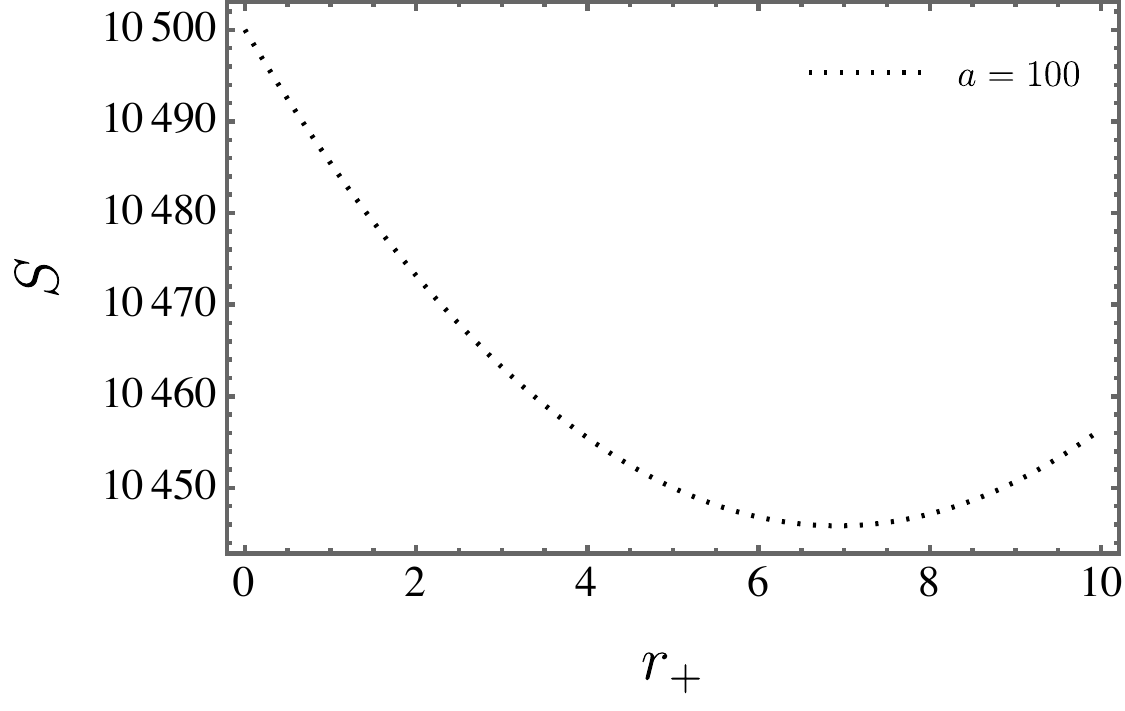}
     \includegraphics[scale=0.4]{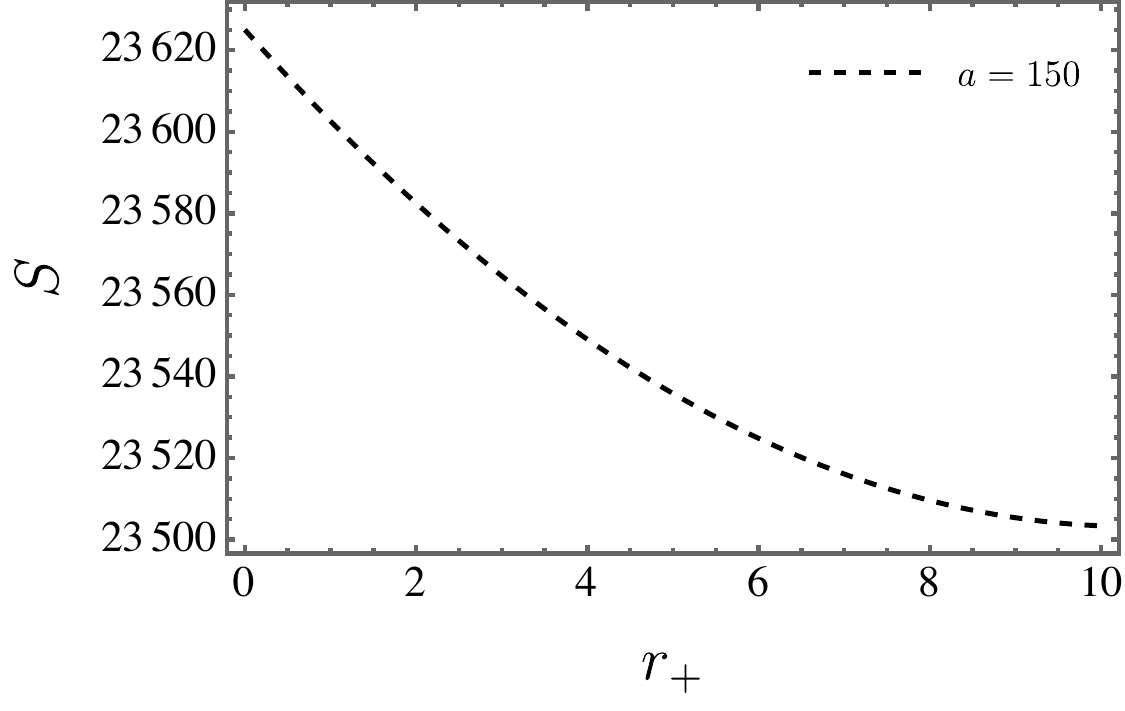}
    \caption{Entropy for different values of $a$, varying the event horizon when $X=0.2$.}
    \label{entropyrotations}
\end{figure}

Since we are interested in obtaining the entropy for our system, now we shall focus on the event horizon area. To formally determine the area of the event horizon, it is essential to consider the following generalized volume element:
\ie
\mathrm{d}V = \sqrt{|g|}\mathrm{d}^{n}x
\fe
where $n$ is the dimension of the manifold. When one considers a hypersurface of Eq. (\ref{metric3}), with the parameters set as $t = \text{constant}$ and $r = r_{\pm}$, it is observed that the generalized volume solely depends on the $\theta$ and $\phi$ coordinates. Thereby, we write
\ie
\begin{split}
 A  = & \int_{0}^{\pi} \int_{0}^{2\pi} \sqrt{|g_{\theta\theta} g_{\phi\phi}|} \, \mathrm{d}\theta \, \mathrm{d}\phi \\
& = -\left[\frac{\pi  \sqrt{\left(a^2+r_{+}^2\right)^2}}{a^2}
\Bigg(X \left(a^2+r_{+}^2\right) \sqrt{\frac{a^2 r_{+}^2}
{\left(a^2+r_{+}^2\right)^2}} \cos ^{-1}\left(\sqrt{\frac{r_{+}^2}
{a^2+r_{+}^2}}\right)\right. \\
& \left. +a \left(r_{+} X \tan ^{-1}
\left(\frac{a}{r_{+}}\right)
 -a (X+4)\right)\Bigg)\right],
\end{split}
\fe
where $X$ is assumed to be small and $A$ is the area calculated by considering the outer horizon only. The entropy associated with this area is expressed as $S = A/4\pi$. To better illustrate this thermodynamic quantity, we present Figs. \ref{entropycomparison} and \ref{entropyrotations}. In Fig. \ref{entropycomparison}, we compare the entropy of a standard Kerr black hole with our case, where we set the parameters to $a = 10$ and $X = 0.2$. Notably, the inclusion of LV effects may lead to at least one potential phase transition. Conversely, in Fig. \ref{entropyrotations}, we investigate how the entropy, with $X = 0.2$, varies as we change the rotation of the black hole, specifically through the parameter $a$.

\subsubsection{The heat capacity}

To conclude the thermodynamic analysis, we compute the heat capacity
\ie
\begin{split}
C_{V} = & T \frac{\mathrm{d}S}{\mathrm{d}T}  = \frac{(a-r_{+}) (a+r_{+}) \sqrt{\left(a^2+r_{+}^2\right)^2}}{4 a^2 \left(a^4+4 a^2 r_{+}^2-r_{+}^4\right)} \\
& \times \left[ X \left(a^2+3 r_{+}^2\right) \left(\left(a^2+r_{+}^2\right) \sqrt{\frac{a^2 r_{+}^2}{\left(a^2+r_{+}^2\right)^2}} \cos ^{-1}\left(\sqrt{\frac{r^2}{a^2+r_{+}^2}}\right)+a r_{+} \tan ^{-1}\left(\frac{a}{r_{+}}\right)\right) \right. \\
& \left. -4 a^2 r_{+}^2 (X+2) \right].
\end{split}
\fe

\begin{figure}
    \centering
     \includegraphics[scale=0.4]{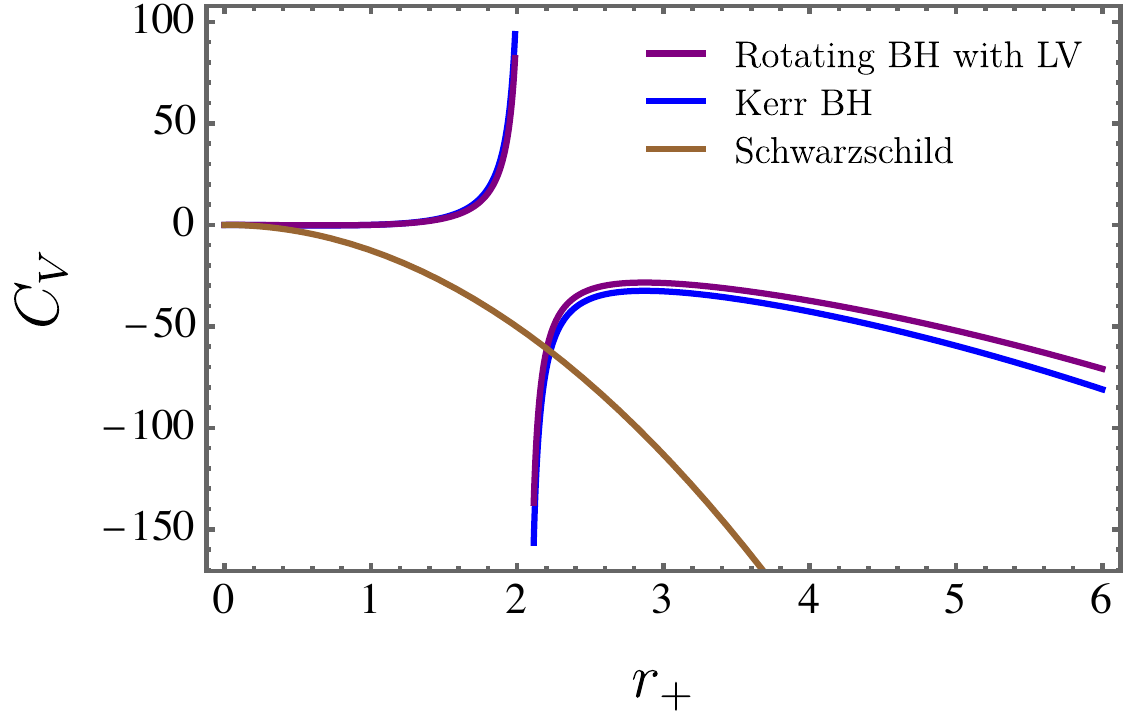}
     \includegraphics[scale=0.4]{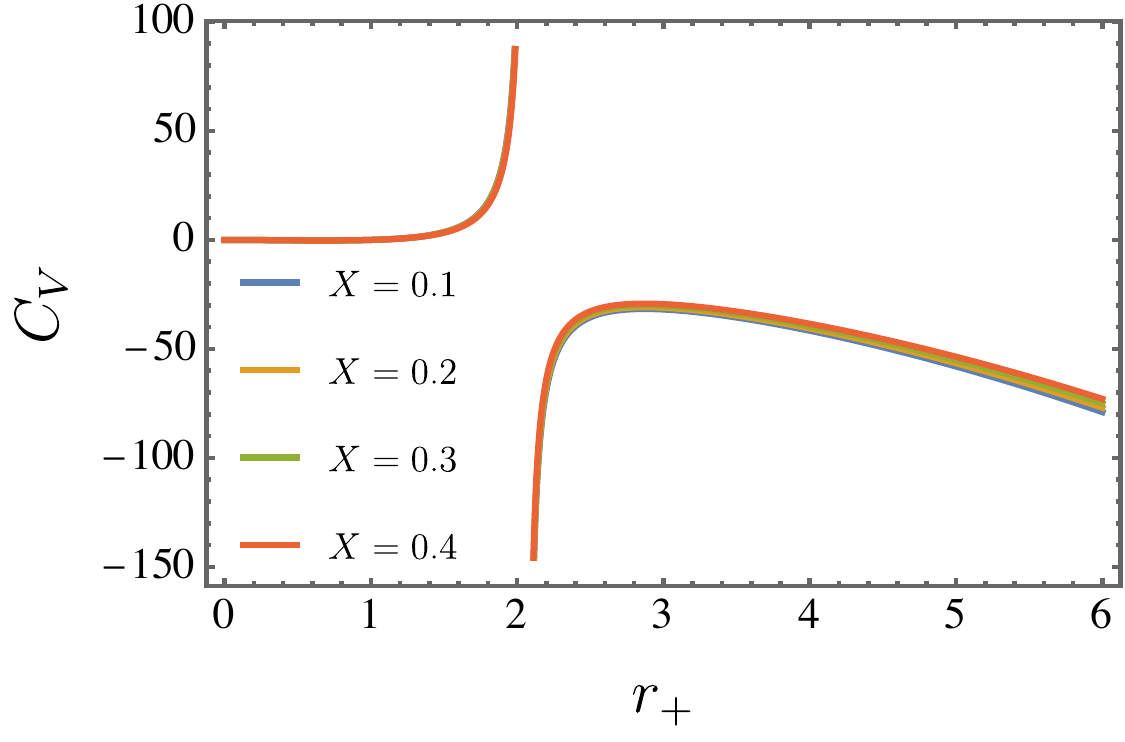}
    \caption{On the left hand, we compare the heat capacity of two black hole solutions with our case. In contrast, on the right hand, we compute $C_{V}$ for different values of $X$. }
    \label{heatcapacity}
\end{figure}

In Fig. \ref{heatcapacity}, we display the behavior of the heat capacity as a function of the horizon $r_{+}$. On the left hand, we show $C_{V}$ for the rotating black hole with LSB (for $X=0.5$), Kerr black, and the Schwarzschild black hole. Notice that in contrast to the latter case, our results indicate a particular region where the stability gives rise to; and phase transitions are also indicated \cite{araujo2023thermodynamical,araujo2022thermal,sedaghatnia2023thermodynamical}. The same feature is also exhibited when Kerr black hole is taken into account. On the other hand, we also present $C_{V}$ for different values of $X$ (on the right hand). As one could expect, there is no expressive modification in the heat capacity, since the parameter $X$ is small.

\subsection{Geodesics}

 Our next objective is to explore the impact of LSB on the geodesic paths of particles moving within the spacetime described by Eq. (\ref{metric3}). Given our axisymmetric metric, it is endowed with two associated Killing vectors: $\partial_{t}$ and $\partial_{\phi}$. This renders it adequate to focus on the radial geodesics. To derive the geodesic equations for point particles, we initiate our analysis with the following Lagrangian, as presented in \cite{Wald}
\ie
\mathcal{L} = g_{\mu\nu}\dot{x}^{\mu}\dot{x}^{\nu}.
\fe
The quantity $\mathcal{L}$ can assume values of $-1, 0, 1$, indicating timelike, null, and spacelike geodesics, respectively. In the previous equation, the dot denotes a derivative with respect to an affine parameter denoted as $\lambda$. We define the velocity as $\dot{x}^{\mu}\equiv \frac{dx^{\mu}}{d\lambda}$. In this sense, we write
\begin{eqnarray}
\mathcal{L} &=&-\left(\frac{\Delta-a^2\sin^2{\theta}}{\rho^2}\right)\frac{\Dot{t}^2}{\sqrt{\left(1+\frac{3X}{4}\right)\left(1-\frac{X}{4}\right)}}\\
\nonumber &-&\frac{4aMr\sin^2{\theta}}{\sqrt{\left(1+\frac{3X}{4}\right)\left(1-\frac{X}{4}\right)}\rho^2}\Dot{t} \Dot{\phi}\\  
   \nonumber &+&\frac{1}{\Delta \sqrt{\left(1+\frac{3X}{4}\right)\left(1-\frac{X}{4}\right)}}\left(a^2 \cos^2 \theta+r^2 \dfrac{\left(1+\frac{3 X}{4}\right)}{\left(1-\frac{X}{4}\right)}\right)\Dot{r}^2 \\
   \nonumber &+&\frac{1}{\sqrt{\left(1+\frac{3X}{4}\right)\left(1-\frac{X}{4}\right)}}\left(r^2 + a^2 \cos^2 {\theta} \dfrac{\left(1+\frac{3 X}{4}\right)}{\left(1-\frac{X}{4}\right)} \right) \Dot{\theta}^2
    \\
    \nonumber &+&\frac{(r^2 +a^2)^2-a^2 \Delta\sin^2{\theta}}{\sqrt{\left(1+\frac{3X}{4}\right)\left(1-\frac{X}{4}\right)}\rho^2}\sin^2 \theta \Dot{\phi}^2 +\\
    \nonumber &+&\frac{2r X a\cos{\theta} }{\sqrt{\left(1+\frac{3X}{4}\right)}\left(1-\frac{X}{4}\right)^{\frac{3}{2}}}\frac{\Dot{r}\Dot{\theta}}{\sqrt{\Delta}}.
    \label{metricr}
\end{eqnarray}
To perform our analysis, we confine the particle motion to the equatorial plane, where $\theta =\frac{\pi}{2}$. Under this restriction, for the metric (\ref{metric3}), we find that:
\begin{eqnarray}
\mathcal{L} &=&-\left(\frac{\Delta-a^2}{r^2}\right)\frac{\Dot{t}^2}{\sqrt{\left(1+\frac{3X}{4}\right)\left(1-\frac{X}{4}\right)}}\\
\nonumber &-&\frac{4aM}{\sqrt{\left(1+\frac{3X}{4}\right)\left(1-\frac{X}{4}\right)}r}\Dot{t} \Dot{\phi}\\  
   \nonumber &+&\frac{r^2}{\Delta } \dfrac{\left(1+\frac{3 X}{4}\right)^{\frac{1}{2}}}{\left(1-\frac{X}{4}\right)^{\frac{3}{2}}}\Dot{r}^2 \\
    \nonumber &+&\frac{(r^2 +a^2)^2-a^2 \Delta}{\sqrt{\left(1+\frac{3X}{4}\right)\left(1-\frac{X}{4}\right)}r^2}\Dot{\phi}^2. 
    \label{metricf}
\end{eqnarray}
Since we have two conserved quantities, $E$ (energy) and $L$ (angular momentum), we can see that
\ie
\label{energy}
E = - g_{t\mu}\dot{x}^{\mu} =  \left(\frac{\Delta-a^{2}}{r^2 \sqrt{\left(1+\frac{3X}{4}\right)\left(1-\frac{X}{4}\right)}}\right) \Dot{t} + \frac{4aM}{\sqrt{\left(1+\frac{3X}{4}\right)\left(1-\frac{X}{4}\right)}r} \Dot{\phi},
\fe
and
\ie
\label{angularmomentum}
L = g_{\phi \mu}\dot{x}^{\mu} = - \frac{4aM}{\sqrt{\left(1+\frac{3X}{4}\right)\left(1-\frac{X}{4}\right)}r} \Dot{t} + \frac{(r^2 +a^2)^2-a^2 \Delta}{\sqrt{\left(1+\frac{3X}{4}\right)\left(1-\frac{X}{4}\right)}r^2} \Dot{\phi}.
\fe
To solve both Eqs. (\ref{energy}) and (\ref{angularmomentum}), let us define them using a shorthand notation, i.e.,
\ie
E = A \Dot{t} + B \Dot{\phi},
\fe
and
\ie
L = - B \Dot{t} + C \Dot{\phi},
\fe
where $A \equiv \frac{\Delta-a^{2}}{r^2 \sqrt{\left(1+\frac{3X}{4}\right)\left(1-\frac{X}{4}\right)}}$, $B \equiv \frac{2aM}{\sqrt{\left(1+\frac{3X}{4}\right)\left(1-\frac{X}{4}\right)}r}$, and $C \equiv  \frac{(r^2 +a^2)^2-a^2 \Delta}{\sqrt{\left(1+\frac{3X}{4}\right)\left(1-\frac{X}{4}\right)}r^2}$. Notice that
\ie
CE - BL = (AC + B^{2})\Dot{t} = \frac{\Delta}{ \left(1+\frac{3X}{4}\right)\left(1-\frac{X}{4}\right)} \Dot{t}
\fe
and
\ie
AL+BE= (AC + B^{2})\Dot{\phi} =  \frac{\Delta}{ \left(1+\frac{3X}{4}\right)\left(1-\frac{X}{4}\right)} \Dot{\phi}.
\fe
where, $ AC + B^{2} = \frac{\Delta}{\left(1+\frac{3X}{4}\right)\left(1-\frac{X}{4}\right)}$.
Therefore, we have
\ie
\Dot{t} = \frac{\left(1+\frac{3X}{4}\right)\left(1-\frac{X}{4}\right)}{\Delta} \left[ \left( \frac{(r^2 +a^2)^2-a^2 \Delta}{\sqrt{\left(1+\frac{3X}{4}\right)\left(1-\frac{X}{4}\right)}r^2} \right)E - \left( \frac{2aM}{\sqrt{\left(1+\frac{3X}{4}\right)\left(1-\frac{X}{4}\right)}r} \right)L  \right],
\fe
\ie
\Dot{\phi} = \frac{\left(1+\frac{3X}{4}\right)\left(1-\frac{X}{4}\right)}{\Delta} \left[  \left( \frac{\Delta-a^2}{r^2 \sqrt{\left(1+\frac{3X}{4}\right)\left(1-\frac{X}{4}\right)}} \right) L + \left( \frac{2aM}{\sqrt{\left(1+\frac{3X}{4}\right)\left(1-\frac{X}{4}\right)}r} \right) E  \right].
\fe
Now, we shall derive the equation governing the radial component of the four-velocity in terms of the variables $A$, $B$, and $C$
\ie
\begin{split}
g_{\mu\nu}\dot{x}^{\mu}\dot{x}^{\nu} & = \mathcal{L} \\
= & - A \Dot{t}^{2} - 2 B \Dot{t} \Dot{\phi} + C \Dot{\phi}^{2} + D \Dot{r}^{2}\\
& = - [A \Dot{t} + B \Dot{\phi}] \Dot{t} + [-B \Dot{t} + C \Dot{\phi}] \Dot{\phi} + D \Dot{r}^{2}\\
& = - E \Dot{t} + L\Dot{\phi} + \frac{D}{\Delta} \Dot{r}^{2},
\end{split}
\fe
where $D = r^2 \dfrac{\left(1+\frac{3 X}{4}\right)^{\frac{1}{2}}}{\left(1-\frac{X}{4}\right)^{\frac{3}{2}}}$. Therefore, the radial equation is given by
\ie
\begin{split}
\Dot{r}^{2} & = \frac{\Delta}{D} \left( E \Dot{t} - L \Dot{\phi} + \mathcal{L} \right) \\
& = \frac{\left(1+\frac{3X}{4}\right)\left(1-\frac{X}{4}\right)}{D} \left[  CE^{2} - 2 BLE -AL^{2}  + \frac{\mathcal{L}\Delta}{\left(1+\frac{3X}{4}\right)\left(1-\frac{X}{4}\right)}  \right] .
\end{split}
\fe
Notice that
\ie  
CE^{2} - 2 BLE -AL^{2}  + \frac{\mathcal{L}\Delta}{\left(1+\frac{3X}{4}\right)\left(1-\frac{X}{4}\right)} = \left( E - \mathcal{V}_{-} \right)\left( E + \mathcal{V}_{+} \right),
\fe
where $\mathcal{V}_{\pm} = \left(\sqrt{4 A C L^2+4 B^2 L^2-\frac{4 C \Delta  \mathcal{L}}{\left(1+\frac{3X}{4}\right)\left(1-\frac{X}{4}\right)}}+2 B L\right)/2 C$, which leads to the radial equation below
\ie
\label{radialequation}
\Dot{r}^{2} = \frac{\left(1+\frac{3X}{4}\right)\left(1-\frac{X}{4}\right)}{D}\left[ \left( E - \mathcal{V}_{-} \right)\left( E + \mathcal{V}_{+} \right) \right].
\fe

Explicitly, ${\mathcal{V}}_{\pm}$ is given by
\ie
\begin{split}
& \mathcal{V}_{\pm} = \frac{ 1}{a^2 (2 M+r)+r^3} \left( \pm r \sqrt{-((X-4) (3 X+4))}  \right)\\
& \times \left[ \sqrt{\left(a^2+r (r-2 M)\right)}\right.\\
&\left.\times \sqrt{-\frac{ \left(4 \mathcal{L} \sqrt{-((      X -4) (3 X+4))} \left(a^2 (2 M+r)+r^3\right)+L^2 r (X-4) (3 X+4)\right)}{r (X-4)^2 (3 X+4)^2}} \right. \\
& \left. +2 a L M \right] .
\end{split}
\fe

Notice that when $X \rightarrow 0$, the potential ascribed to Kerr black hole is recovered as one should expect. Also, it is worth mentioning that, if we define the quantity $\overset{\nsim}{\mathcal{V}} \equiv \mathcal{V}_{+} \cdot \mathcal{V}_{-} = \left[ L^2 (r-2 M)-\frac{4 r \mathcal{L} \left(a^2+r (r-2 M)\right)}{\sqrt{4-X} \sqrt{3 X+4}}\right]/ \left[ a^2 (2 M+r)+r^3\right]$ and consider $X \rightarrow 0$, and $a \rightarrow 0$, the usual effective potential to the Schwarzschild case is recovered.

\begin{figure}
    \centering
     \includegraphics[scale=0.4]{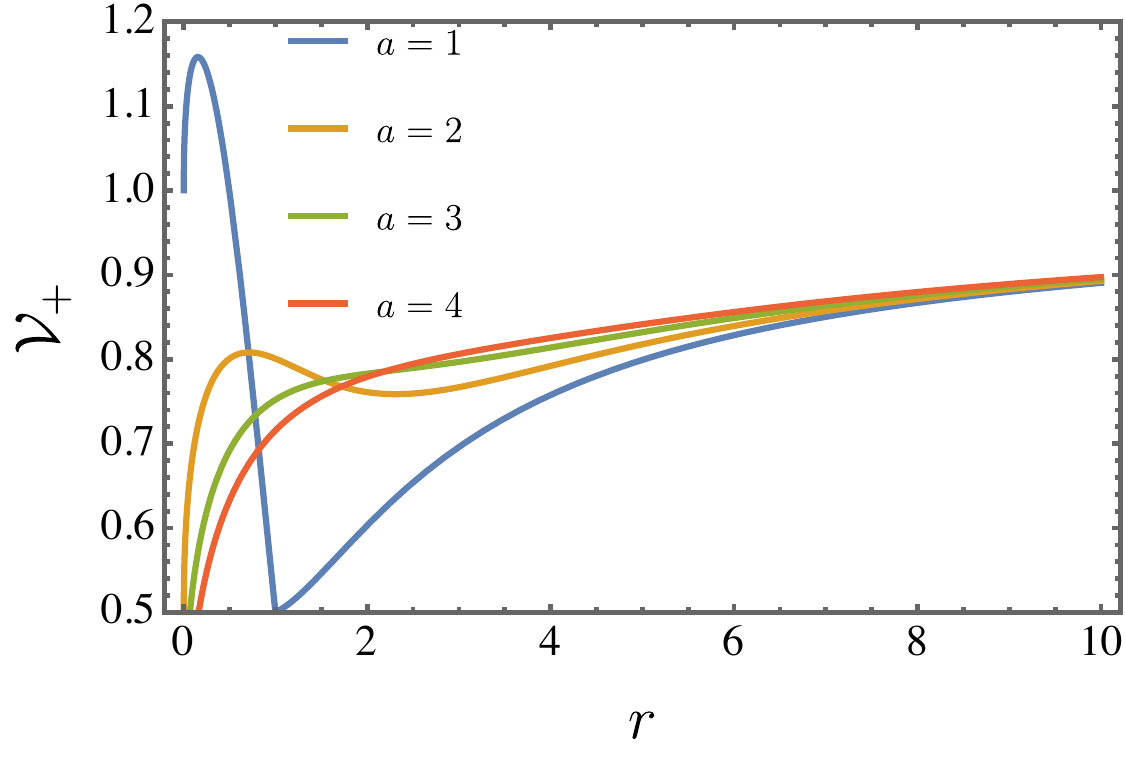}
    \includegraphics[scale=0.41]{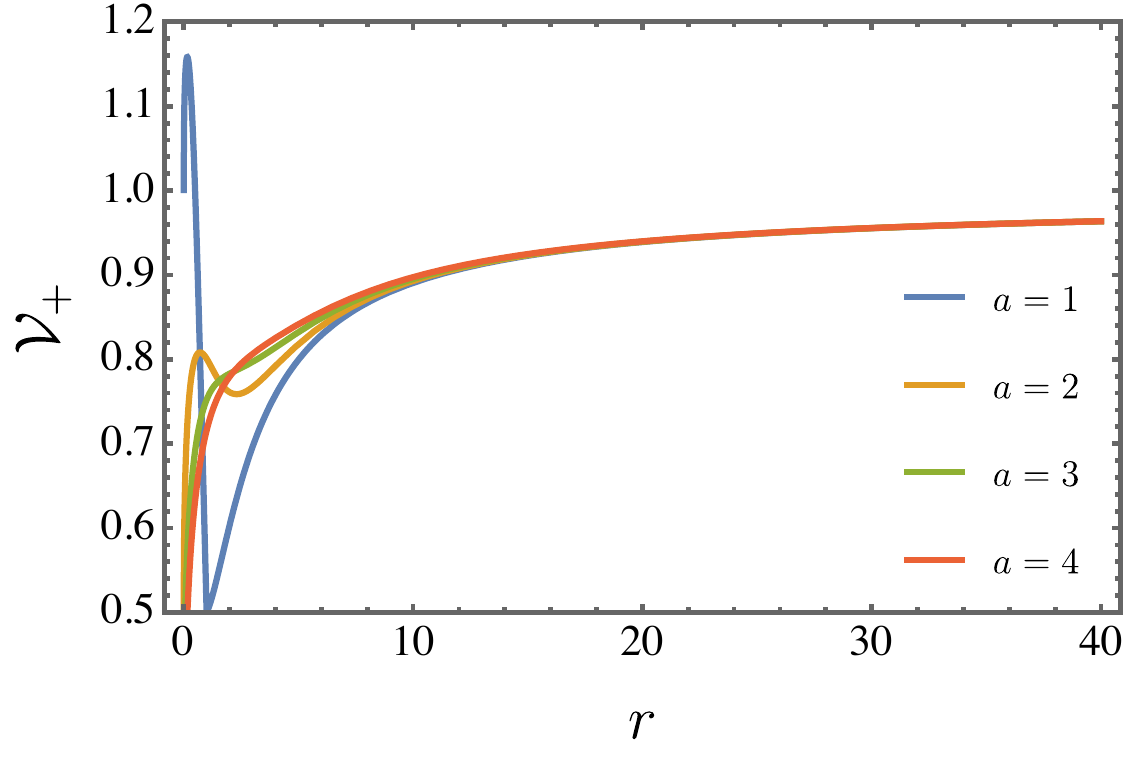}
    \includegraphics[scale=0.41]{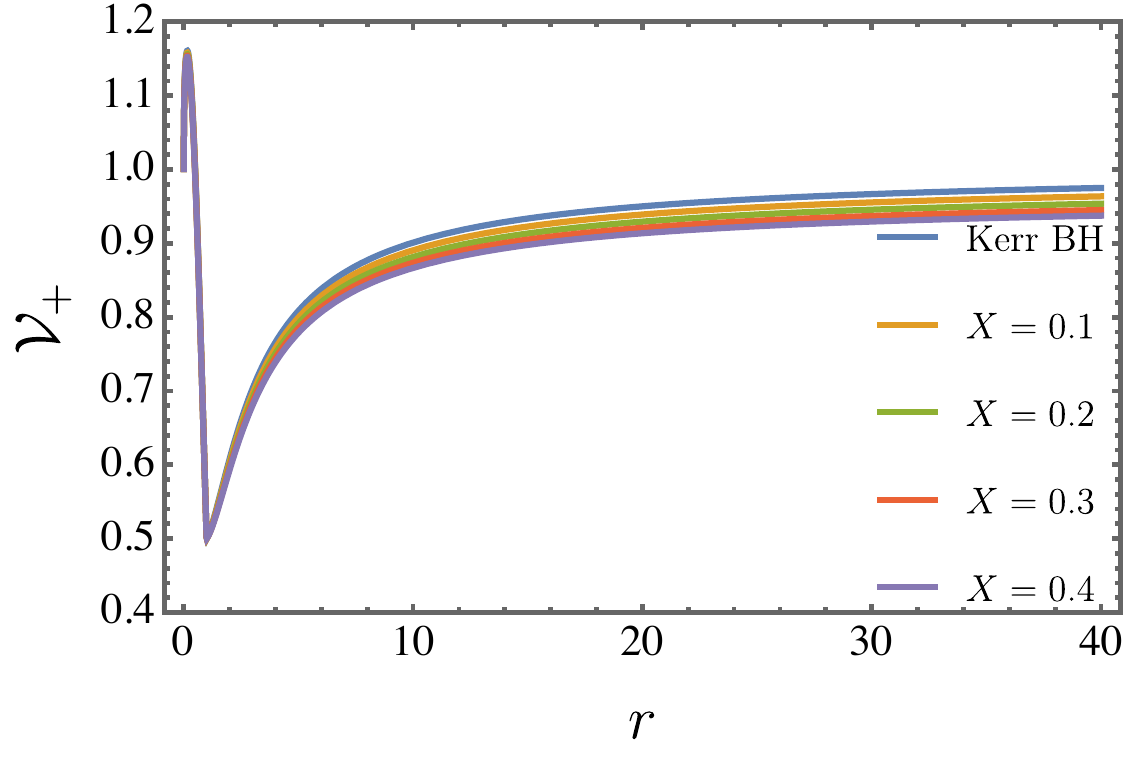}
    \caption{Potential $\mathcal{V}_{+}$ is represented for different configurations of $a$ and $X$.}
    \label{potentialsplus}
\end{figure}

\begin{figure}
    \centering
     \includegraphics[scale=0.4]{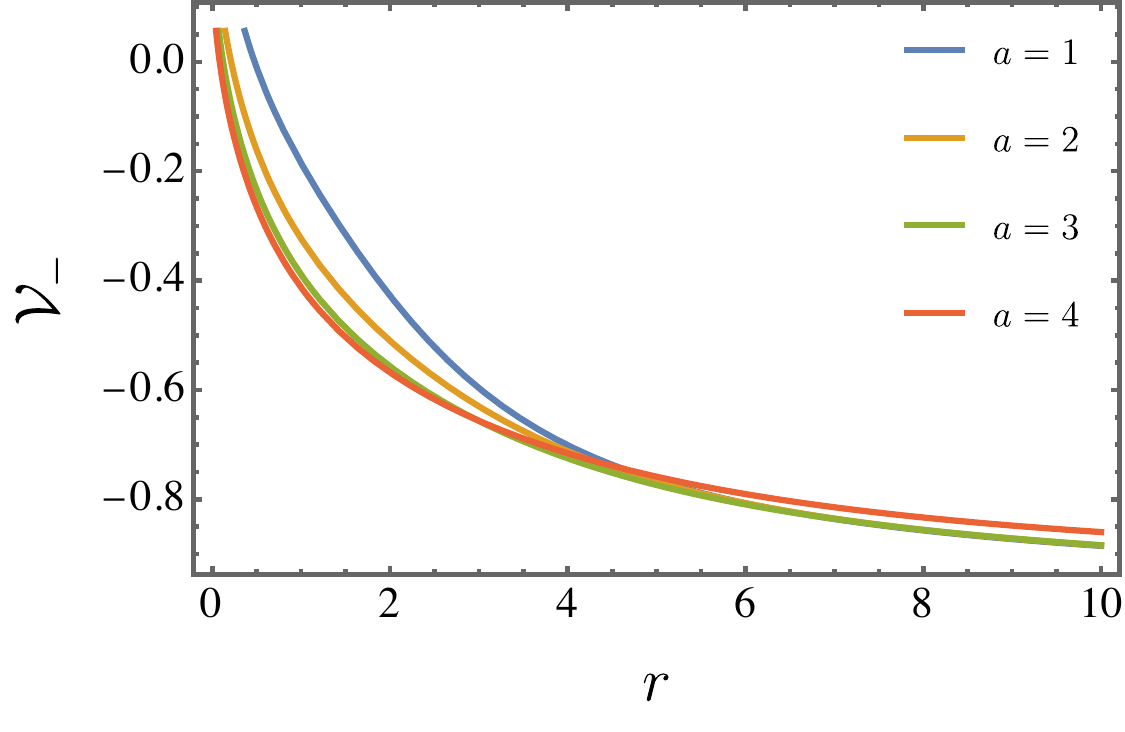}
    \includegraphics[scale=0.41]{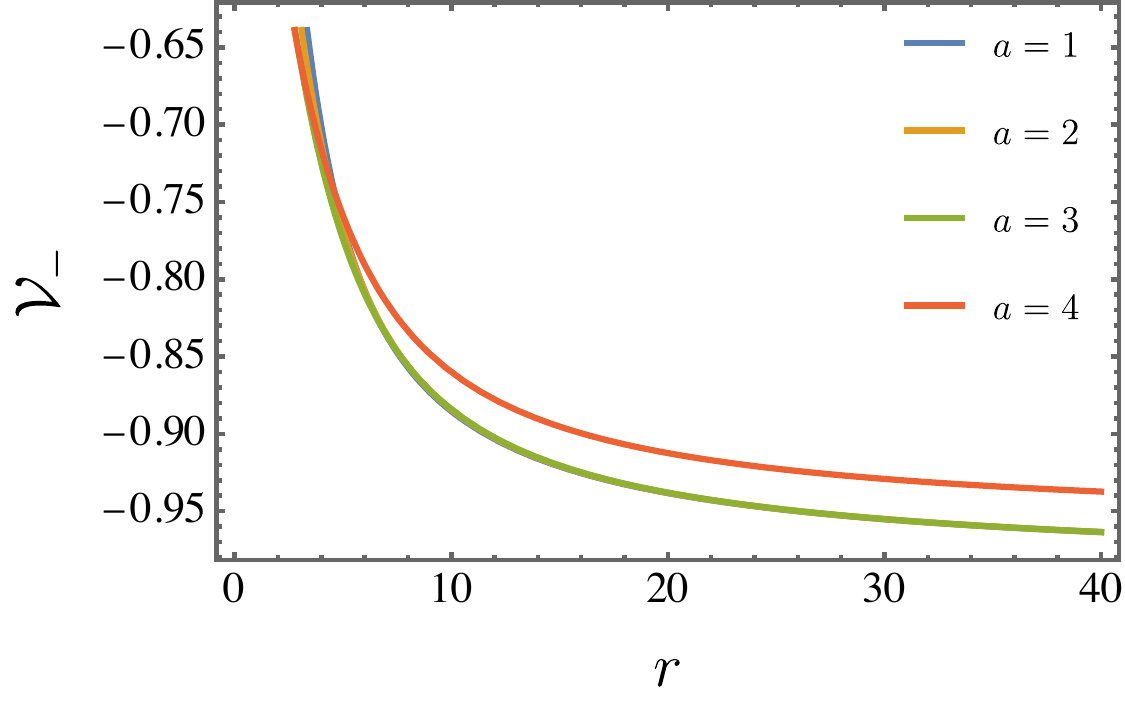}
    \includegraphics[scale=0.41]{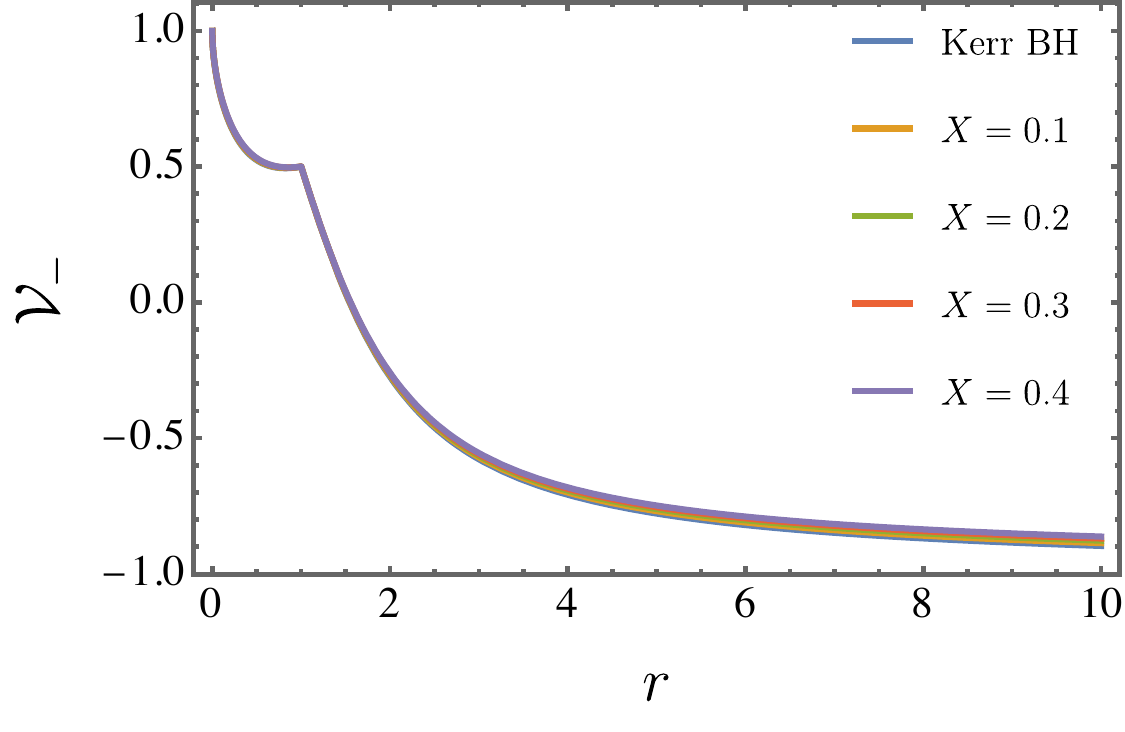}
    \caption{Potential $\mathcal{V}_{-}$ is represented for for different configurations of $a$ and $X$.}
    \label{potentialsminus}
\end{figure}

To demonstrate the characteristics of the potentials \(\mathcal{V}_{\pm}\), Figs. \ref{potentialsplus} and \ref{potentialsminus} are presented concerning the timelike configuration, i.e., $\mathcal{L} = -1$. In Fig. \ref{potentialsplus}, we present the behavior of $\mathcal{V}_{+}$ as a function of $r$. The figure displays diverse values of $a$ on the left and right, while maintaining a fixed value of $X$. Additionally, different values of $X$ are depicted for a fixed $a$ at the bottom, facilitating a comparison with the Kerr black hole. In addition, Fig. \ref{potentialsminus}, a similar analysis is also accomplished, i.e., using the same values of $X$ and $a$, considering $\mathcal{V}_{-}$ though.

Here, we could calculate the critical orbits by considering $\frac{\mathrm{d} \mathcal{V}_{\pm}}{\mathrm{d} r} = 0$. Nevertheless, we verified that $X$ plays no role in the photon sphere. The same feature has been recently reported in the literature for the spherically symmetric case \cite{hassanabadi2023gravitational}.

\subsubsection{The radial acceleration analysis for null geodesics}

In the context of null geodesics, the radial Eq. (\ref{radialequation}) reads 
\ie
\begin{split}
\Dot{r}^{2} & =  \frac{\left(1+\frac{3X}{4}\right)\left(1-\frac{X}{4}\right)}{D} \left( E - \mathcal{V}_{-} \right)\left( E + \mathcal{V}_{+} \right)  \\
& = \frac{(4-X)^{5/2} \sqrt{3 X+4}}{64 r^2} \left( E - \mathcal{V}_{-} \right)\left( E + \mathcal{V}_{+} \right).
\end{split}
\fe
As the square of $\Dot{r}^{2}$ must be positive, examination of the above expression reveals that null geodesics are viable for a massless particle, being $(4-X)^{5/2} \sqrt{3 X+4} > 0$, when the constant of motion $E$ satisfies the inequalities dictated by
\ie
E < \mathcal{V}_{-} \,\, \,\, \text{or} \,\,\,\, E > \mathcal{V}_{+}\,\,.
\fe
In other words, $\mathcal{V}_{-} < E < \mathcal{V}_{+}$ turns out to be a forbidden region. In addition, to analyze the orbits effectively, it is beneficial to calculate the radial acceleration. By differentiating equation (\ref{radialequation}) with respect to the parameter $s$, we obtain:
\ie
\begin{split}
2 \dot{r} \ddot{r} = & \left[ \left( \frac{\left(1+\frac{3X}{4}\right)\left(1-\frac{X}{4}\right)}{D} \right)^{\prime} (E- \mathcal{V}_{+})(E-\mathcal{V}_{-}) - \frac{\left(1+\frac{3X}{4}\right)\left(1-\frac{X}{4}\right)}{D}\mathcal{V}_{+}^{\prime}(E-\mathcal{V}_{-})\right.\\
 & \left.  - \frac{\left(1+\frac{3X}{4}\right)\left(1-\frac{X}{4}\right)}{D}\mathcal{V}_{-}^{\prime}(E-\mathcal{V}_{+}) \right] \Dot{r},
\end{split}
\fe
or
\ie
\begin{split}
\Ddot{r} = & \frac{1}{2}\left( \frac{\left(1+\frac{3X}{4}\right)\left(1-\frac{X}{4}\right)}{D} \right)^{\prime} (E- \mathcal{V}_{+})(E-\mathcal{V}_{-}) \\ 
&- \frac{\left(1+\frac{3X}{4}\right)\left(1-\frac{X}{4}\right)}{2 D} \left[ \mathcal{V}_{+}^{\prime}(E-\mathcal{V}_{-}) - \mathcal{V}_{-}^{\prime}(E-\mathcal{V}_{+}) \right].
\end{split}
\fe
Here, the prime symbol ``$\prime$'' denotes differentiation with respect to $r$. Now, let us systematically examine the radial acceleration at a point where the radial velocity, $\dot{r}$, is equal to zero, i.e., specifically when the energy parameter \( E \) equals the potential energy $\mathcal{V}_{+}$ or $\mathcal{V}_{-}$:
\ie
 \Ddot{r} = - \frac{\left(1+\frac{3X}{4}\right)\left(1-\frac{X}{4}\right)}{2 D}  \mathcal{V}_{+}^{\prime}(\mathcal{V}_{+} -\mathcal{V}_{-}), \,\,\,\,\,\,\text{if} \,\,\,\,\, E = \mathcal{V}_{+},
\fe
and
\ie
 \Ddot{r} = - \frac{\left(1+\frac{3X}{4}\right)\left(1-\frac{X}{4}\right)}{2 D}  \mathcal{V}_{-}^{\prime}(\mathcal{V}_{-}-\mathcal{V}_{+}), \,\,\,\,\,\,\text{if} \,\,\,\,\, E = \mathcal{V}_{-}.
\fe
Moreover, since
\ie
\mathcal{V}_{+} - \mathcal{V}_{-} = \frac{2 r \sqrt{-((X-4) (3 X+4))} \sqrt{-\frac{L^2 \left(a^2+r (r-2 M)\right)}{(X-4) (3 X+4)}}}{a^2 (2 M+r)+r^3},
\fe
we obtain
\ie
 \Ddot{r}_{\pm} = \mp \frac{\left(\frac{3 X}{4}+1\right) \left(1-\frac{X}{4}\right)}{2 D} \left( \frac{2 r \sqrt{-((X-4) (3 X+4))} \sqrt{-\frac{L^2 \left(a^2+r (r-2 M)\right)}{(X-4) (3 X+4)}}}{a^2 (2 M+r)+r^3} \right) \mathcal{V}'_{\pm}, \,\,\,\, \text{if} \,\,\,\, E = \mathcal{V}_{\pm}.
\fe

\begin{figure}
    \centering
     \includegraphics[scale=0.4]{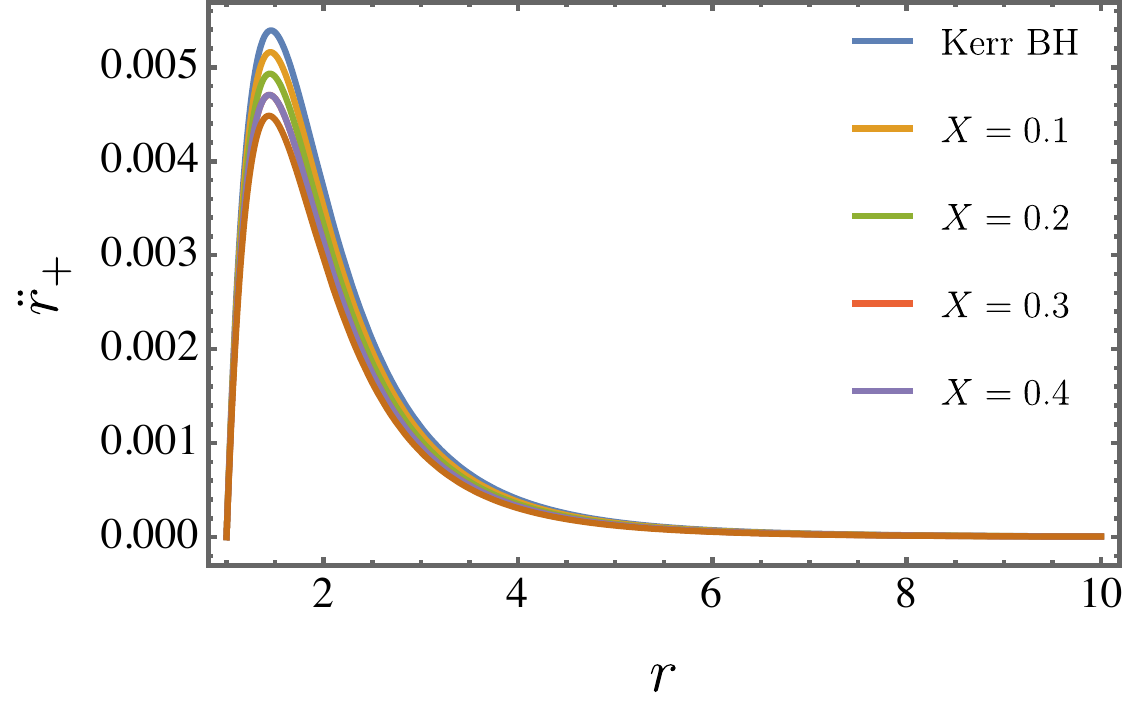}
    \includegraphics[scale=0.405]{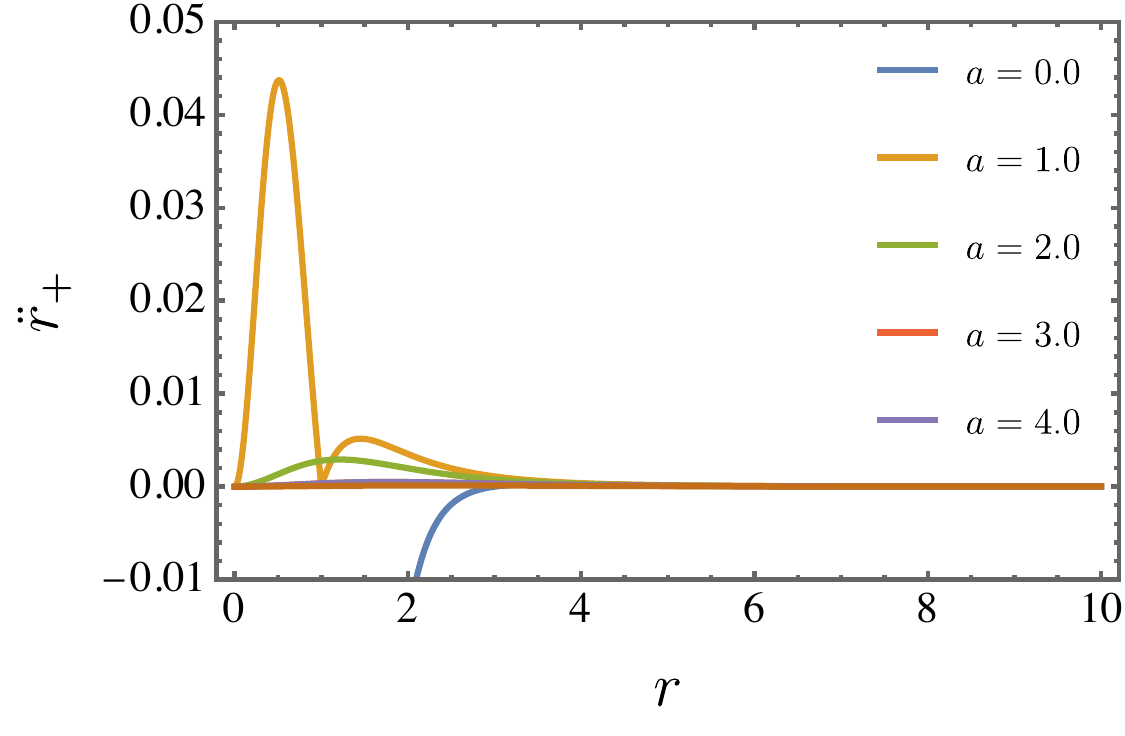}
    \caption{Acceleration $\Ddot{r}_{+}$  is represented as a function of $r$ for different configurations of $a$ and $X$.}
    \label{rplusa}
\end{figure}

\begin{figure}
    \centering
     \includegraphics[scale=0.4]{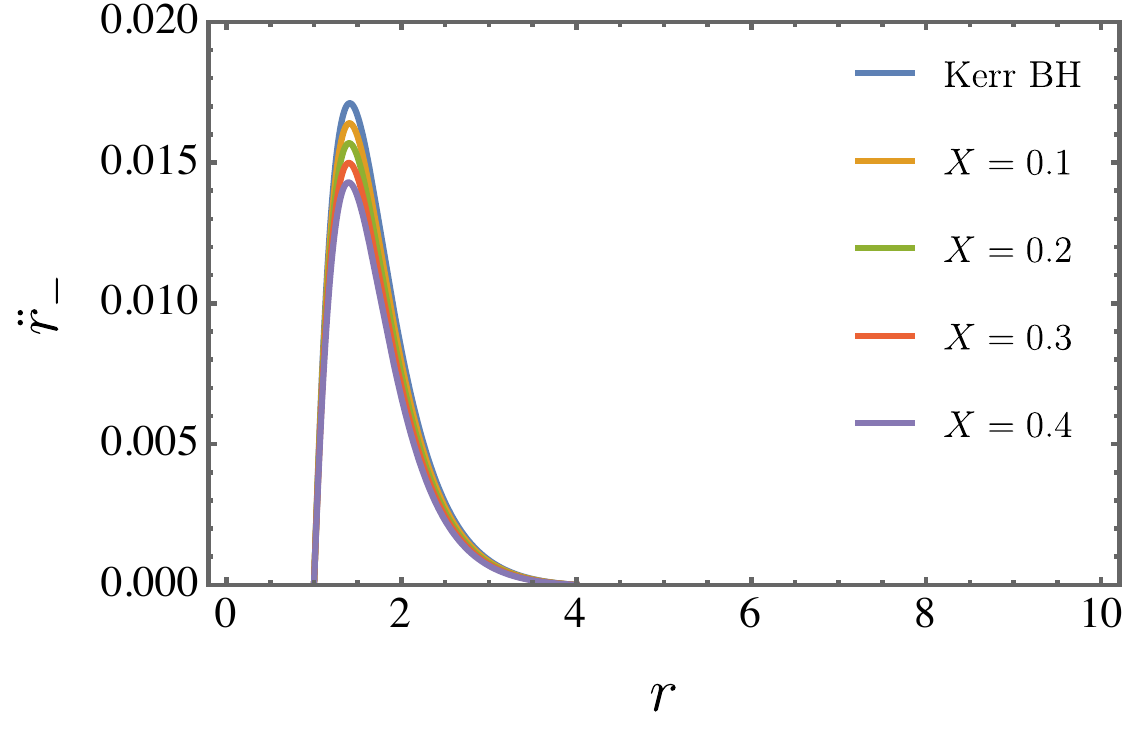}
    \includegraphics[scale=0.39]{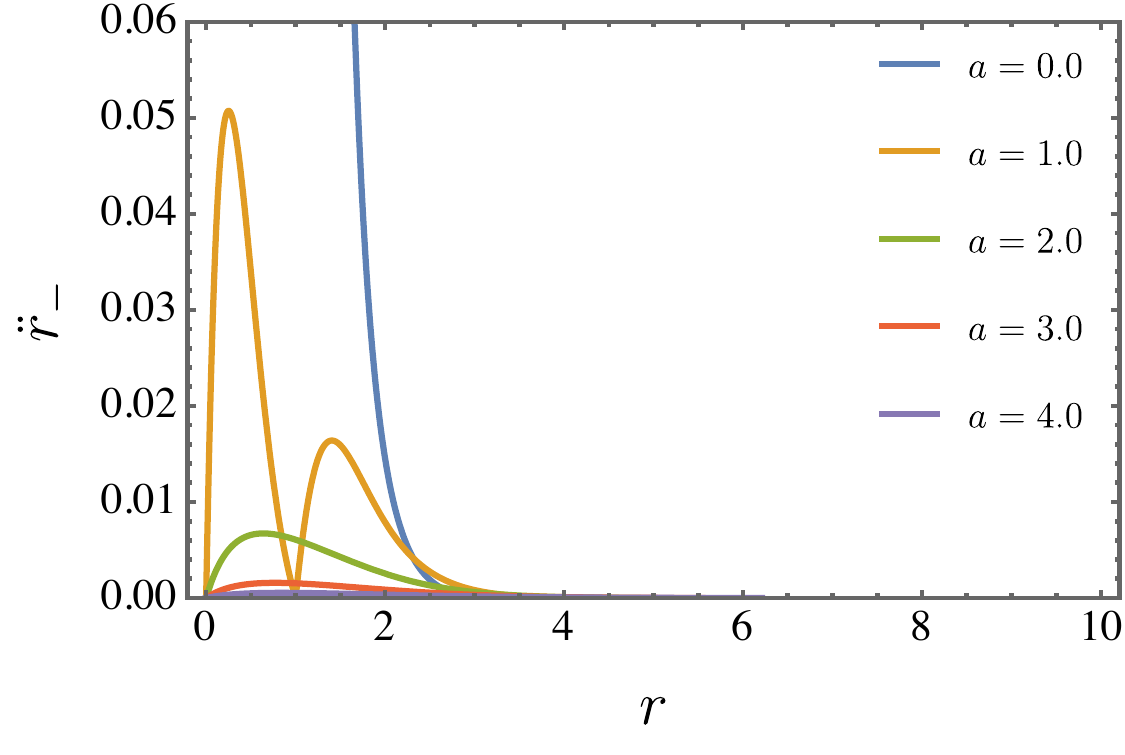}
    \caption{Acceleration $\Ddot{r}_{-}$  is represented as a function of $r$ for different configurations of $a$ and $X$.}
    \label{rminusa}
\end{figure}

For the sake of a better interpretation of $\Ddot{r}_{\pm}$, we provide Figs. \ref{rplusa} and \ref{rminusa}.

\subsubsection{Time-like geodesics and the advance of Mercury’s perihelion in the LSB Kerr-like spacetime}

We systematically investigate the repercussions of LSB on the geodesics of both massive and massless test particles in the innermost regions. This departure is anticipated to diverge from the established behavior dictated by GR. Here, the time-like geodesics read
\ie
\nonumber
\Dot{r}^{2} = \frac{\left(\frac{3 X}{4}+1\right) \left(1-\frac{X}{4}\right)}{D}\left[ \left( E - \mathcal{V}_{-} \right)\left( E + \mathcal{V}_{+} \right) \right].
\fe

We investigate the impact of LSB corrections on the Innermost Stable Orbit (ISCO), a crucial scenario where two circular orbits approach and merge. Our goal is to discern these corrections from the predictions of GR. Initial examinations reveal that solving the equation \( \frac{\mathrm{d}\mathcal{V}_{\pm}}{\mathrm{d}r} = 0 \) as a function of \( X \) indicates that LSB does not alter the ISCO. Nevertheless, it is anticipated that noncircular orbits will be influenced by LSB. To illustrate, we focus on the precession of Mercury's perihelion. As a well-established phenomenon, the first step involves expressing the radial coordinate \( r \) in terms of angular variables \( \phi \), denoted as \( r(\phi) \) 
\ie
\begin{split}
\left( \frac{\mathrm{d}r}{\mathrm{d}\phi} \right)^{2} = &  -\frac{1}{\left.r (3 X+4)^2 (2 a E M+L (r-2 M))^2\right)} \\
& \times \left[  \left(a^2+r (r-2 M)\right)^2 \left(E^2 (X-4) (3 X+4) \left(a^2 (2 M+r)+r^3\right) \right.\right. \\
& \left.\left. +4 a^2 r \sqrt{-((X-4) (3 X+4))}-4 a E L M (X-4) (3 X+4) \right.\right. \\
& \left.\left. +L^2 (X-4) (3 X+4) (2 M-r)-8 M r^2 \sqrt{-((X-4) (3 X+4))} \right.\right. \\
& \left. \left. +4 r^3 \sqrt{-((X-4) (3 X+4))}\right) \right].
\end{split}
\fe

Now, let us consider $r=L^{2}/M y$, which follows
\ie
\begin{split}
& \left( \frac{\mathrm{d}y}{\mathrm{d}\phi} \right)^{2} = - \frac{1}{L^6 M^2 (3 X+4)^2 \left(2 a E M^2 y+L^3-2 L M^2 y\right)^2} \\
& \times \left\{ \left(a^2 M^2 y^2+L^4-2 L^2 M^2 y\right)^2 \left[ 4 L^6 \sqrt{-((X-4) (3 X+4))} \right.\right.\\
& \left.\left. -8 L^4 M^2 \sqrt{-((X-4) (3 X+4))} y +4 a^2 L^2 M^2 \sqrt{-((X-4) (3 X+4))} y^2 \right.\right. \\
& \left.\left.   L^6 (X-4) (3 X+4)E^2     +2 M^4 (X-4) (3 X+4) y^3 (L-a E)^2   \right.\right. \\
& \left. \left.  -L^2 M^2 (X-4) (3 X+4) y^2 (L-a E) (a E+L)    \right]       \right\}.
\label{differential}
\end{split}
\fe
Notice that, if we consider $X = 0$ and $a = 0$, we recover the differential equation which governs the Schwarzschild metric. The next crucial step involves transforming it into a second-order differential equation. This is accomplished by taking the derivative of Eq. (\ref{differential}) with respect to $\phi$ and performing additional straightforward algebraic manipulations. Furthermore, we shall restrict our analysis to the slow rotation approximation, which corresponds to considering the dimensionless rotation parameter $\frac{a}{M}<<1$, and also $X<<1$, as usual. By doing so, Eq.\eqref{differential} becomes
\ie
\begin{split}
0&=\frac{\mathrm{d}^2 y}{\mathrm{d}\phi^2}-\frac{ 3 M^2}{ L^2}y(\phi )^2 +y(\phi ) \left(\frac{8 a E^3  M^2}{L^3}-X+1\right)+C
\end{split}
\label{geode1}
\fe
where $C = \frac{2aE}{L}(E^{2}-1) -1 -\frac{5}{4}X$. The modifications due to LV corrections do not exert influence on the first and second terms in the \textit{r.h.s} of the aforementioned equation. While their impact is restricted to the remaining other terms, as can be seen in the previous equation. When the LV coefficients are set to be zero, the slow rotation limit of the Kerr solution is recovered, as expected. The most effective approach to discern their effects is through a perturbative treatment of Eq. (\ref{geode1}). Consequently, the solution can be elegantly expressed in the following form:
\ie
\label{perttt}
y = y_{0} + y_{1} + ... \,\,\,.
\fe
Here, $y_{0}$ represents the unperturbed case, which corresponds to the Newtonian solution corrected by the LV coefficient; while $y_{1}$ denotes the first-order perturbed solution, which takes into account the contributions stemming from the slow rotation and LV parameters $a$ and $X$, respectively. It is noteworthy that, within the perturbative framework employed here, higher-order corrections will be disregarded. In this sense, substituting  Eq. (\ref{perttt}) into Eq. (\ref{geode1}) and solving the resulting equation iteratively, we obtain the zeroth-order solution
\ie
y_{0} = 1 + \epsilon \cos\phi.
\fe
where $\epsilon$ is the eccentricity. The former equation is precisely the Newtonian result. For the first-order solution, we have
\ie
\begin{split}
y_{1} = & 3\gamma\left(1+\frac{2}{3}\epsilon^2-\frac{8aE^3}{3L}\right)-\frac{2a E}{L}(E^2 -1)-\frac{1}{4}X-\gamma\epsilon^2 \cos^2{\phi}+\\
&+\left(\frac{1}{2}X+3\gamma-\frac{4aE^3 \gamma}{L}\right)\epsilon\cos{\phi}+\left(3\gamma+\frac{1}{2}X-\frac{4aE^3 \gamma}{L}\right)\epsilon\phi\sin{\phi},
\label{eqpert}
\end{split}
\fe
where we have defined the constant quantity  $\gamma=\frac{M^2}{L^2}$. In practical terms, note that only the last term in Eq.\eqref{eqpert} plays a non-trivial role since the other ones represent constants and oscillatory terms around zero. Therefore, the relevant physical piece of the solution is
\begin{equation}
    y= 1 + \epsilon \cos\phi+\epsilon\phi  \sin\phi  \left( 3\gamma-4\frac{E^3 a}{L}\gamma+\frac{1}{2}X\right).
\end{equation}
From the observational data $\gamma<<1$ for Mercury \cite{d1992introducing} and by the fact that $X$ must also be much smaller than $1$, then the previous equation can be set into the following form 
\begin{equation}
    y=1+\epsilon\cos\left((1-3\gamma+4\frac{E^3 a}{L}\gamma-\frac{1}{2}X)\phi\right),
    \label{name}
\end{equation}
up to first-order in $X$ and $\gamma$.
The perihelion shift $(\Delta\beta)$ is computed from Eq.\eqref{name} by defining the period of non-circular orbits 
\begin{equation}
    T_{0}=\frac{2\pi}{(1-3\gamma+4\frac{E^3 a}{L}\gamma-\frac{1}{2}X)}\approx 2\pi +\Delta\beta,
\end{equation}
where the advance of perihelion contribution can be split into three pieces, namely,
\begin{equation}
    \Delta\beta=6\pi\gamma +\delta_{_{Kerr}}\beta+\delta_{_{LV}}\beta.
\end{equation}
The first one is the usual Schwarzschild contribution
\begin{equation}
    \Delta\beta_{0}=6\pi\gamma= \frac{6\pi GM}{c^2(1-\epsilon^2)\tilde{a}},
\end{equation}
where we have made use of the identification $L^2=\frac{GM}{c^2}(1-\epsilon^2)\tilde{a}$, with $\tilde{a}$ denoting the semi-major axis of the orbital ellipse, and restored the Newton's constant $G$ and the speed of light $c$. The second contribution takes into account the first-order effects of rotation coming from the Kerr metric,
\begin{equation}
    \delta_{_{Kerr}}\beta=-\frac{8\pi G^2 M^2 E^3 a}{c^4 L^3}\approx -\frac{8\pi G^2 M^2 a}{c L^3},
    \label{kerrb}
\end{equation}
where we used the approximation $E\approx c$ above. Such a contribution measures the impact of the Sun's rotation on the Mercury's perihelion. Finally, the third contribution is entirely due to the LSB,
\begin{equation}
    \delta_{_{LV}}\beta=\pi X .
    \label{LVb}
\end{equation}
The contributions (\ref{kerrb}) and (\ref{LVb}) should be seen as corrections to the standard Schwarzschild contribution for the advance of the perihelion.

\subsubsection{Estimation of the LSB coefficient from the advance of Mercury’s perihelion}

In order to estimate the LSB coefficient, one can use the GR theoretical prediction for the advance of Mercury's perihelion and then compare it with the astrophysics data at disposal. Therefore, by using the theoretical data \cite{Link, Iorio:2018adf}, the standard contribution is well known $\Delta\beta_{0}=42.981^{\prime\prime}/\mbox{century}$, while the Kerr contribution is given by
\begin{equation}
\delta_{_{Kerr}}\beta=-0.002^{\prime\prime}/\mbox{century}.
\label{kerrk}
\end{equation}
Note that this result is in accordance with the observational data obtained from planetary Lense-Thirring precession \cite{er, er1}. In addition, recent observational data show a correction of the order $-0.002\pm 0.003 ^{\prime\prime}/\mbox{century}$ for the Mercury's perihelion \cite{Pit, Pit1}, which reveals that \eqref{kerrk} is within the experimental error. On the other hand, it is well known that LSB effects have not already been observed by current experiments, thereby one might estimate an upper bound for the LV coefficient $X$. This methodology consists of inferring that the contribution of the LSB for the Mercury's perihelion should not be bigger than the observational uncertainty, ($0.003^{\prime\prime}/\mbox{century}$ or $72.3\times 10^{-7\,\prime\prime}/\mbox{orbit}$), i.e.,
\begin{equation}
    \delta_{_{LV}}\beta<0.003^{\prime\prime}/\mbox{century}=72.3\times 10^{-7\,\prime\prime}/\mbox{orbit}.
\end{equation}
By doing so, we are able to estimate the upper bound to be $X<4.9\times 10^{-12}$, which is in agreement with that estimation found in the previous work \cite{Filho:2022yrk} for the Schwarzschild-like metric with LSB.

\section{Summary and conclusion}
\label{summary}

Finding exact rotating solutions within the modified theories of gravity is more than a challenging task. Despite the complexities inherent in deriving these solutions due to the involved structure of the modified field equations, the search for exact rotating solutions in modified theories of gravity stands as a prominent program to probe the strong gravitational field regime, giving us valuable insights on new physics beyond GR.

 In this work, in order to find a new exact rotating solution, we considered a particular modified theory of gravity called the metric-affine traceless bumblebee gravity, i.e., a theory known to be ghost-free in its gravitational sector. We also derived the respective field equations for the dynamical fields: the bumblebee field, the metric and the connection. Regarding the latter quantity, we found the Levi-Civita connection of the auxiliary metric $h_{\mu\nu}$ as the solution, in which $h-$ and $g-$metrics were related to each other by a disformal transformation \eqref{metric2}. It allowed us to entirely rewrite the modified Einstein field equations in terms of $h_{\mu\nu}$ (Einstein frame), which is the natural metric that incorporates the effects of LSB.

Knowing the vacuum gravitational field equations in the Einstein frame, we began with the Kerr metric as a seed, which is a consistent solution of such equations in the Einstein frame, to find the solution of the physical metric $g_{\mu\nu}$. In addition, we assumed that the bumblebee field stays frozen in the VEV and, different from \cite{Filho:2022yrk}, it possesses a non-trivial angular dependency. Keeping all this information in mind, we found that the $g-$metric described a stationary and axisymmetric solution with LSB codified by the parameter $X=\xi b^2$. This new solution presented some remarkable features. Firstly, the new gravitational corrections to the Kerr (GR) solution emerged as non-linear terms in $X$. Secondly, we recovered the usual Kerr solution as $X\to 0$. Finally, when $a\to 0$, the same solution was reduced to the Schwarzschild-like one \cite{Filho:2022yrk} after a suitable rescaling. 

Having the novel solution, we then explored the impact of the LSB on the thermodynamics properties. We have calculated the \textit{Hawking} temperature, the entropy, and the heat capacity. We showed that the LV parameter does not affect the event horizon and Cauchy horizon radii. However, the entropy and heat capacity were affected by the LV parameter, which departed from the standard phase transition curves as $X$ grows.
In addition, we obtained the time-like and null geodesics and showed that they were affected by corrections in $X$, as expected. To conclude, we provided estimations for the LV parameter by confronting the theoretical predictions with the available astrophysical data of the advance of Mercury's perihelion. In this case, we found an upper bound limit for the LV coefficient  $X<4.9\times 10^{-12}$, which was in agreement with that one found in \cite{Filho:2022yrk}.

To provide a more comprehensive insight into our study, we intend to investigate further aspects of our novel modified Kerr-like black hole. This includes exploring its potential impact within the context of gravitational lenses, quasinormal modes, and other relevant issues. These and other ideas are now under development.

\section*{Acknowledgments}
\hspace{0.5cm}The authors would like to thank the Conselho Nacional de Desenvolvimento Cient\'{\i}fico e Tecnol\'{o}gico (CNPq) for financial support. P. J. Porf\'{\i}rio would like to acknowledge the Brazilian agency CNPQ, grant No. 307628/2022-1. The work by A. Yu. Petrov. has been partially supported by the CNPq project No. 303777/2023-0. Moreover, A. A. Araújo Filho is supported by Conselho Nacional de Desenvolvimento Cient\'{\i}fico e Tecnol\'{o}gico (CNPq) and Fundação de Apoio à Pesquisa do Estado da Paraíba (FAPESQ) -- [150891/2023-7].

\bibliographystyle{ieeetr}
\bibliography{main}

\end{document}